\documentclass[english,aps,longbibliography,twocolumn]{revtex4-1}
\usepackage[T1]{fontenc}
\usepackage[latin9]{inputenc}
\usepackage{amsmath}
\usepackage{amssymb}
\usepackage{graphicx}
\usepackage{babel}
\setcitestyle{super}

\begin{document}

\title{Wavevector multiplexed atomic quantum memory \\via spatially-resolved single-photon detection}

\author{M. Parniak\textsuperscript{1,\textdagger}, M. D\k{a}browski\textsuperscript{1,*},
M. Mazelanik\textsuperscript{1}, A. Leszczy\'{n}ski\textsuperscript{1},
M. Lipka\textsuperscript{1}, and W. Wasilewski\textsuperscript{1}}

\affiliation{\textsuperscript{1}Institute of Experimental Physics, Faculty of
Physics, University of Warsaw, Pasteura 5, 02-093 Warsaw, Poland}

\begin{abstract}
Parallelized quantum information processing requires tailored quantum memories to simultaneously handle multiple photons. The spatial degree of freedom is a promising candidate to facilitate such photonic multiplexing. 
Using a single-photon resolving camera we demonstrate a wavevector multiplexed quantum memory based on a cold atomic ensemble. Observation of nonclassical 
correlations between Raman scattered photons is confirmed by an average value of the second-order correlation function $g_{\mathrm{{S,AS}}}^{(2)}=72\pm5$ in 665 separated modes simultaneously. 
The proposed protocol utilizing the multimode
memory along with the camera will facilitate generation of multi-photon
states, which are a necessity in quantum-enhanced sensing technologies and as an input to photonic quantum circuits.

\end{abstract}
\maketitle

\section*{Introduction}

Multiplexing in optical fibers or in free-space is essential in modern techniques for high-throughput
transmission \cite{Wang2012,Richardson2013}. Similarly, as quantum technologies
mature, the necessity of multiplexing in photon-based quantum communication
becomes clear \cite{Munro2010} and much effort is devoted to various schemes exploiting
spatial \cite{Surmacz2008,Lan2009,Dai2012a,Ding2013,Nicolas2014,Parigi2015,Lee2016,Chen2016,Dabrowski2017}
temporal  \cite{Clausen2011,Collins2013,Humphreys2014,Gundogan2015,Cho2016,Xiong2016} or spectral \cite{Sinclair2014,Puigibert2017}
degrees of freedom.
Utilization of many modes can finally allow efficient
application of the Duan-Lukin-Cirac-Zoller (DLCZ) protocol \cite{Duan2001,Simon2007,Sinclair2014,Kutluer2017,Laplane2017}
and offer nearly-deterministic generation of multi-photon states \cite{Ma2011,Nunn2013,Chrapkiewicz2017}
later applicable in quantum enhanced sensing technologies \cite{Wolfgramm2012,Matthews2016} as well as optical quantum computation \cite{Kok2007}.

Regardless of substantial efforts, the task of achieving large number
of modes remains a challenging endeavor especially in hybrid atom-photon
systems. 
In purely photonic, memoryless systems
hundreds of modes have been obtained within the spatial domain of
spontaneous parametric down-conversion (SPDC) \cite{Edgar2012,Moreau2014,Krenn2014}
or by means of frequency-time entanglement \cite{Roslund2013,Kaneda2015,Xie2015,Reimer2016,Xiong2016,Cai2017}.
However, most applications such as the DLCZ protocol \cite{Duan2001},
enhanced photon generation \cite{Nunn2013,Chrapkiewicz2017,Kaneda2017}
or even linear optical quantum computing (LOQC) \cite{Kok2007} require
or greatly benefit from a multimode quantum memory. For instance, the largest number
of temporal modes used for photon storage in cavity-based quantum memory\cite{Kaneda2015} is 30. Likewise, ensembles of dopants in crystals have been used to store externally generated photons\cite{Tang2015} in up to 100 modes or otherwise generate them\cite{Kutluer2017,Laplane2017} in 12 modes.

Another mainstream trend
is to build a multiplexed quantum repeater by splitting a trapped
atomic ensemble into many cells \cite{Lan2009,Nunn2013}. The idea
was recently realized in two dimensions achieving 225 modes \cite{Pu2017}.
These schemes however suffer from the limitation given by difficulty
in trapping large ensembles as well as hinder heralded simultaneous
excitation of all modes. In consequence, they are rendered useful only for the DLCZ
quantum repeater \cite{Duan2001} but neither for quantum imaging \cite{Boyer2008,Brida2010,Genovese2016}
nor enhancing rate of the photonic state generation \cite{Nunn2013,Sinclair2014,Xiong2016,Chrapkiewicz2017}.

The purpose of this paper is a demonstration of
massive improvement in the number of modes processed
by the quantum memory. The experimental realization is accomplished through multiplexing of angular
emission modes of a single quantum memory \cite{Chrapkiewicz2017} and by employing a spatially-resolved
single-photon detection. Our experimental setup generates photons in 665 pairwise-coupled
modes, exploring the regime of multimode capacity
with simultaneous extremely low noise-level achieved with stringent,
spatially-multimode yet simple and robust filtering. We use a single-photon resolving camera to measure both correlations and auto-correlation unambiguously proving quantum character of light. Note that throughout our results no accidental
or noise background subtraction is performed - in contrast
to any previous experiments with single-photon sensitive cameras \cite{Edgar2012,Moreau2014,Chrapkiewicz2017,Dabrowski2017}.
We achieve the quantum memory lifetime of more
than $50\ \mu\mathrm{s}$ which combined with the multimode capacity invites real-time feedback processing of
stored excitations \cite{Mazelanik2016} and paves the way towards promptly achieving fast generation of single and multi-photon states \cite{Nunn2013,Chrapkiewicz2017}.  

\section*{Results}

\begin{figure}[t]
\includegraphics[scale=1.2]{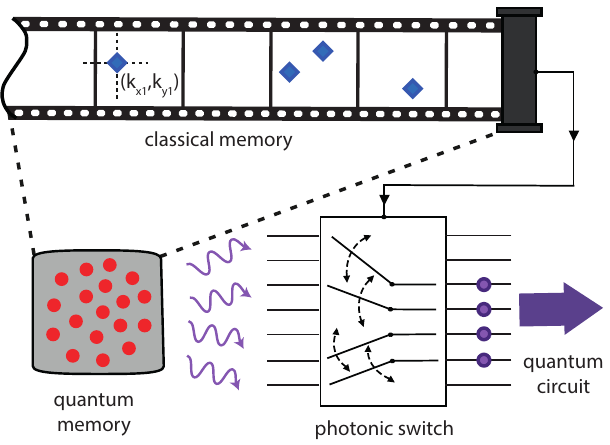}

\caption{\textbf{Proposal for single photon spatial routing and multiplexing for multi-photon
state generation. }A single-photon resolving camera registers consecutive
frames during the quantum memory write-in process. Each detection heralds creation
of a spin-wave excitation in the atomic ensemble quantum memory with a wavevector
determined by the position $(k_{xi},k_{yi})$ at which $i$-th photon was registered. Acquisition
and write-in continue until the desired number of excitations has been
created. At this time a photonic switch is reconfigured to channel
photons from conjugate directions stored in a classical memory and subsequently the readout pulse
is applied to convert stored spin-wave excitations to the requested number of photons, which will be used later, e.g. in the quantum circuit.\label{fig:multiplex}}
\end{figure}

\textbf{Multi-photon generation.}
Here we propose the potential application of our scheme as a platform for multi-photon state generation. Figure~\ref{fig:multiplex} pictures a protocol utilizing the multi-pixel
capability of the single-photon resolving camera to enhance generation
of multi-photon states. 
The essential advantage over recently introduced quantum memory arrays \cite{Pu2017}
is simultaneous excitation and access to many modes. The protocol
is being managed by a classical memory storing the wavevectors of registered
photons and the which-mode information. This information is
finally used to route the photons retrieved
in the readout process from the quantum memory. A photonic switch is used to direct photons to a quantum circuit or to conditionally generate arbitrary states through multi-photon interference. Furthermore, since
a small number of photons is generated per each frame, one can adapt in real-time
the number of trials to create exactly the desired number of excitations
in the quantum memory. By keeping the mean photon number per shot small we virtually eliminate the malicious pairs in a single mode. This gives us an advantage over a simpler scheme\cite{Chrapkiewicz2017} in
which a single excitation shot is used. In that scheme the
mean number of photons may be controlled but the multi-mode thermal
statistics severely limits the fidelity of generation of $n$-photon
state. Extensions of our new proposal are numerous, including usage of spin-wave echos
to conditionally manipulate the atomic excitations \cite{Hosseini2009}.  The experimental results presented below constitute the most essential step towards realization of the proposed protocol.

\begin{figure*}
\includegraphics{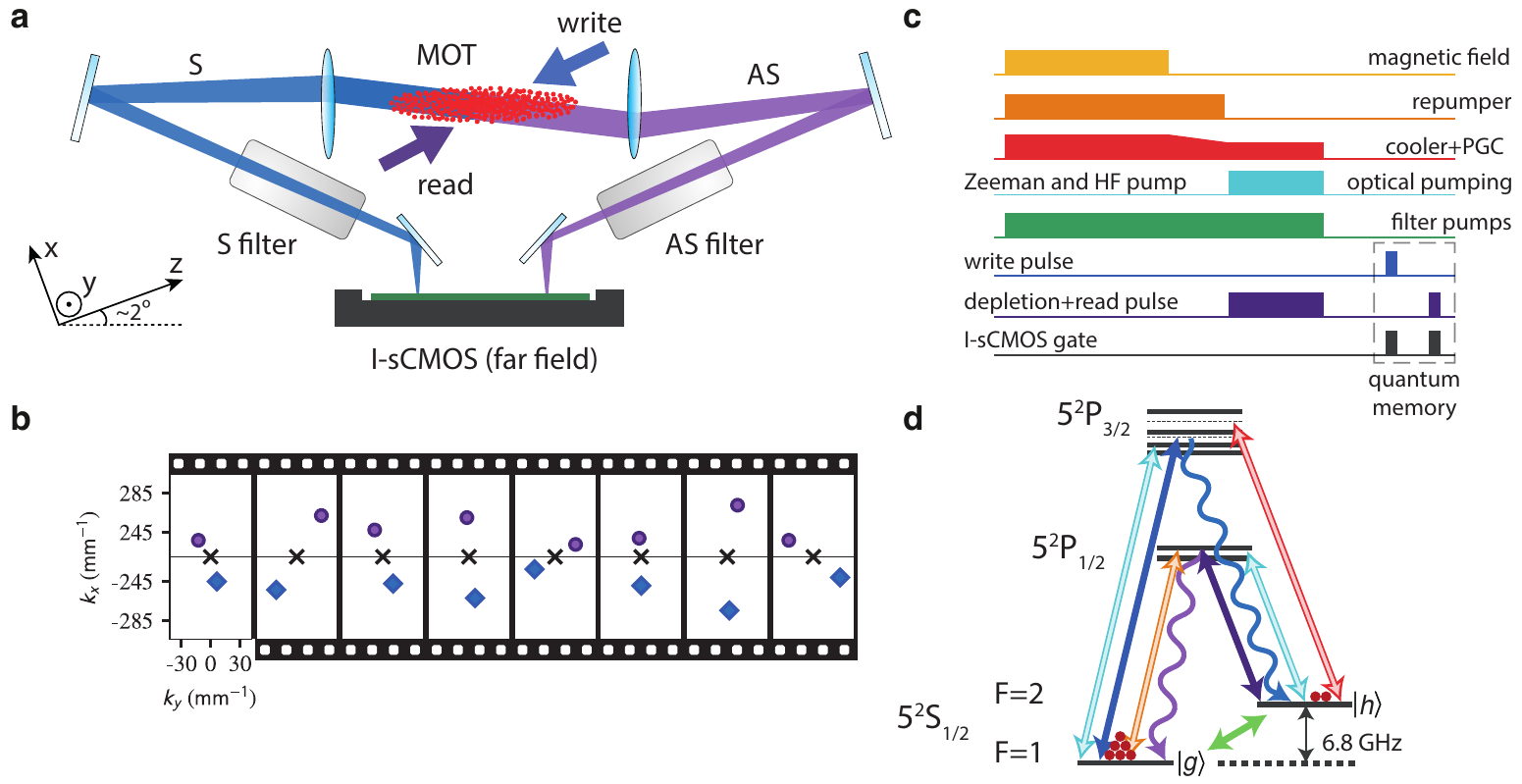}\caption{\textbf{Experimental realization of the wavevector multiplexed quantum memory. (a)}
Schematic of the main part of the experimental setup. The atomic ensemble
released from the magneto-optical trap (MOT) is illuminated with orthogonal
circularly-polarized write and read laser beams. Angles at which Stokes
(S) and anti-Stokes (AS) photons (produced through Raman scattering) are emitted from the atomic ensemble
are imaged on the single-photon resolving I-sCMOS sensor, composed
of a sCMOS camera and an image intensifier. Optically pumped atomic cells
(S and AS filters) filter out the residual laser light and stray fluorescence;
\textbf{(b)} Exemple subsequent frames in the Stokes (bottom) and
anti-Stokes (top) regions demonstrating correlated photon pairs in
each camera frame. Note that while most frames will contain no photon
or a photon only in a single region, almost all (>90\%) frames with
a coincidence event will contain a correlated photon pair for the detection probability of S photon $p_{\mathrm{{S}}}=1.2\times10^{-2}$;
\textbf{(c)} Pulse sequence used in the experiment (see Methods for
details) consists of trapping magnetic field switching, laser cooling
and optical pumping (with depletion) preparation stages, as well as
short write-in and readout laser pulses (quantum memory stage) producing
Stokes and anti-Stokes photons; \textbf{(d)}
Atomic level configuration (colors correspond to the pulse sequence in panel c).
In the process of write-in and readout the spin-wave (green arrow)
is created and annihilated, respectively. The wavy arrows correspond
to S (blue) and AS (purple) photons.\label{fig:setup}}
\end{figure*}

\textbf{Wavevector multiplexing.} For the quantum memory we use an engineered atomic ensemble of cold rubidium-87 atoms at $T=22\pm2\ \mu\mathrm{K}$
generated within a magneto-optical trap (MOT) and cooled using polarization-gradient
cooling (PGC) scheme, as depicted in Fig. \ref{fig:setup}. With the
1 cm-long cigar-shape ensemble of diameter $w=0.6\pm0.1$ $\mathrm{mm}$ (taken as $1/e^{2}$
diameter of the atomic column density) containing $N=10^{8}$
atoms we achieve optical depth $\mathrm{OD}=40$, which limits the memory
readout efficiency \cite{Zhang2012,Cho2016}. Quantum memory operates
once atoms are released from MOT with the magnetic field gradients
switched off. We prepare 70\% of atoms in the  $F=1,\ m_{F}=1$ state and the rest of atoms in the $F=1,\ m_{F}=0$ state through optical pumping.  Atom-photon interface is achieved with two lasers: write, which
is red-detuned from $5^{2}S_{1/2},\ F=1\rightarrow5^{2}P_{3/2},\ F=2$
transition and read laser tuned to $5^{2}S_{1/2},\ F=2\rightarrow5^{2}P_{1/2},\ F=2$
transition. 

To generate the multimode, multi-photon state we illuminate the ensemble
with a write pulse containing $10^{7}$ photons with wavevector $\mathbf{k}_{\mathrm{w}}$
tilted at an angle of $2^{\circ}$ to the ensemble axis in the $xz$
plane. We take the axis defined by counter-propagating read and write
beams as the $z$ axis of the frame of reference. Stokes (S) photons
scattered in the Raman process are registered on the I-sCMOS camera
located in the far-field with respect to the atomic ensemble. A scattered
photon with a transverse wavevector $\mathbf{k}_{\mathrm{S}}$ is
accompanied by a collective atomic excitation (spin-wave) with a spatial
phase dependance: $N^{-1/2}\sum_{j=1}^{N}\exp\left(i\mathbf{K}\cdot\mathbf{r}_{j}\right)|g_{1}\ldots h_{j}\ldots g_{N}\rangle$,
where $\mathbf{K}=\mathbf{k}_{\mathrm{S}}-\mathbf{k}_{\mathrm{w}}$ is the spin-wave wavevector,
$\mathbf{r}_{j}$ is the position of $j$-th atom, $|g_{j}\rangle$ ($|h_{j}\rangle$) corresponds to the $5^{2}S_{1/2},\ F=1$ ($F=2$) state and the summation is
carried over all $N$ atoms in the ensemble. To learn about the spin-wave
we convert it to an anti-Stokes (AS) photon through resonant Raman scattering
(readout process) with a read pulse with wavevector $\mathbf{k}_{\mathrm{r}}$
containing $10^{8}$ photons. Wavevector of the AS photon is determined
by the stored atomic excitation $\mathbf{k}_{\mathrm{AS}}=\mathbf{K}+\mathbf{k}_{\mathrm{r}}$.
We estimate the readout efficiency as $\chi_\mathrm{R}=35\pm2\%$ (taken as the ratio
of coincidence rate to S photons rate and accounted for losses). AS
photons are registered on a separate region of the same I-sCMOS sensor.

Spatially-insensitive filtering is essential for the memory to take
advantage of its inherent multimode capability. Commonly used frequency
filtering cavities \cite{Lan2009,Clausen2011,Chen2016} transmit only
one spatial mode. To overcome this issue we use two separate optically-pumped
hot rubidium vapor cells with buffer gas and paraffin coating. The
cells are pumped by strong lasers during the cooling and trapping
period of the MOT. Additional interference filters are used to separate
stray pump laser light from single photons (see Methods for details).

Finally, photons originating from the atomic quantum memory are imaged
onto the I-sCMOS sensor through a nearly diffraction-limited imaging
setup. The sensor is located in the Fourier plane of the atomic ensemble.
Positions of photons registered on the camera are calibrated as transverse
emission angles, directly proportional to transverse wavevector components.
The I-sCMOS camera has the quantum efficiency of 20\% and the
combined average transmission of the imaging and filtering system
is 40\% (see Methods for details). The net efficiencies in S and AS arms are equal $\eta_{\mathrm{S}}\approx\eta_{\mathrm{AS}}\approx8\%$,

\textbf{Data analysis.}  The spatial degree of freedom provides an advantage over single-mode experiments \cite{Nicolas2014,Parigi2015}.
If one considers each mode as a separate realization
of the experiment we are able to collect statistics at a rate of $3\times10^{5}$
effective experiments per second. This rate is very similar to what
is obtained in single-mode experiments, however the multimode scheme
offers much more versatility as increasing the memory storage time to many
$\mu$s decreases the rate very insignificantly, contrasted with a dramatic drop of the rate in the single-mode experiments. For example, with 30 $\mu$s storage time our effective experimental rate remains at 300 kHz, as it is anyway limited by the readout speed of the sCMOS camera. For the corresponding single-mode experiment the absolute maximum stands at 33~kHz. With faster camera acquisition rate the advantage of the multimode scenario would become overwhelming.  

Here, to obtain proper statistics we have collected $10^{7}$ camera frames. For a pair of small conjugate square-shaped region-of-interests (ROI) with side length $\kappa=160\ \mathrm{mm}^{-1}$ and a net S photon detection probability of $4\times10^{-2}$ we register very few accidental coincidences, i.e. 90\% of coincidences come from conjugate modes. This figure of merit changes with a mean photon number and thus the number of observed modes, as due to limited detection efficiencies we will sometimes register a pair of photons from two different pairwise-coupled modes. For two conjugate ROIs with a side lengths $\kappa=340\ \mathrm{mm}^{-1}$ (i.e. 43 mrad) corresponding to a nearly full field of view composed of hundreds of modes we have registered a total number of $1.6\times10^{5}$ coincidences of which $4.4\times10^{4}$ came from conjugate mode pairs.

\begin{figure}
\includegraphics[scale=0.8]{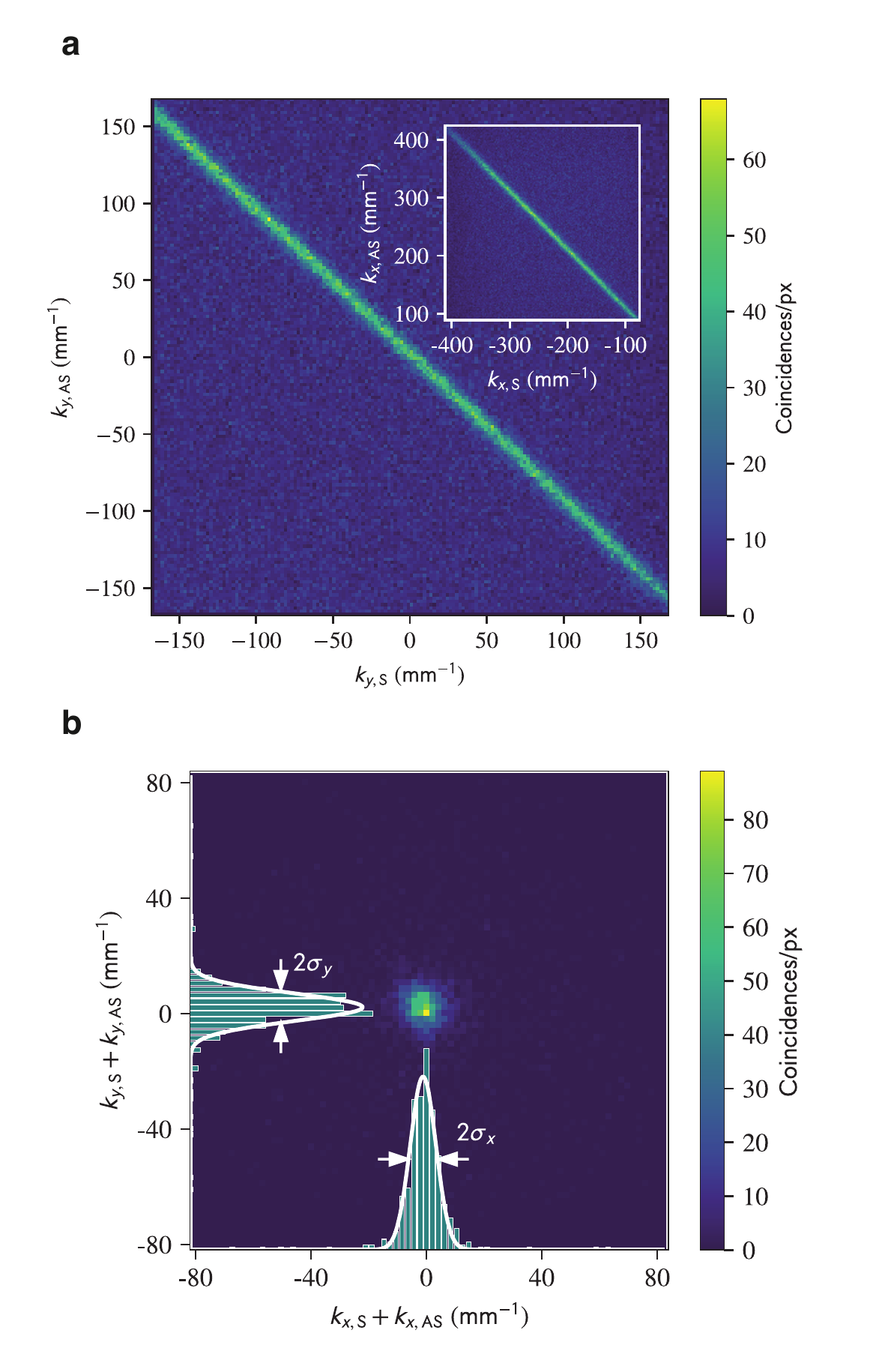}

\caption{\textbf{Spatial properties of the generated biphoton state. (a)} All
Stokes\textendash anti-Stokes coincidences obtained from $10^7$ frames marked with their $k_{y}$
wavevector components $(k_{x}$ for the inset) for zero memory storage time, demonstrating high
degree of momenta anti-correlation. Each plot disregards the perpendicular
component. \textbf{(b)} Same coincidences counted in the center-of-mass
variables $(k_{x,\mathrm{S}}+k_{x,\mathrm{AS}})$ and $(k_{y,\mathrm{S}}+k_{y,\mathrm{AS}})$.
The central peak is fitted with a two-dimensional Gaussian to obtain
its center and width. One-dimensional distributions correspond to
cross-section counts selected for central pixels. Both in (a) and
(b) neither accidental nor noise background subtraction is performed.
\label{fig:mapy}}
\end{figure}

Collection of photon counts with a multi-pixel
detector requires new experimental and data
analysis tools \cite{Chrapkiewicz2014}. To verify the anti-correlation between momenta of S and AS photons both
in $x$ and $y$ coordinates, we count
the coincidences for each pair of camera pixels corresponding to wavevector
coordinates $(k_{x,\mathrm{{S}}},k_{y,\mathrm{S}})\ \mathrm{and}\ (k_{x,\mathrm{AS}},k_{y,\mathrm{AS}})$.
Figure \ref{fig:mapy}a portrays the number of coincidences for a
large field of view as a function of $(k_{y,\mathrm{S}},\ k_{y,\mathrm{AS}})$
momenta summed over the $x$ coordinates. Notably, thanks to a very high signal-to-noise ratio we do not subtract the accidental and noise background in contrast to hitherto schemes\cite{Edgar2012,Moreau2014,Chrapkiewicz2017,Dabrowski2017}.
We observe a clear anti-correlated behavior which we model by the
quantum amplitude of the generated Stokes\textendash anti-Stokes photon
pair in $y$ dimension, given by: 
\begin{equation}
\Psi_y(k_{y,\mathrm{S}},k_{y,\mathrm{AS}})=\mathcal{{N}}\exp\left(-\frac{(k_{y,\mathrm{S}}+k_{y,\mathrm{AS}})^{2}}{2\sigma_{y}^{2}}\right),\label{eq:biampl}
\end{equation}
where $\sigma_{y}$ is a correlation length in the $y$ dimension and $\mathcal{N}$ is a normalization
constant. In turn, the number of coincidences is proportional
to $|\Psi_y(k_{y,\mathrm{S}},k_{y,\mathrm{AS}})|^{2}$. An identical expression describes
photons behavior in $x$ dimension \textendash{} see inset in Fig.
\ref{fig:mapy}a. For the Gaussian-shaped atomic ensemble the size
of the emission mode should be related to ensemble transverse dimension
$w=0.6\pm0.1$ $\mathrm{mm}$, corresponding to wavevector spread
of $2/w=3.3\pm0.5\ \mathrm{{mm}}^{-1}$ in the far-field for light at the wavelength of S photons. To precisely determine the
mode widths $\sigma_{x,y}$, in Fig. \ref{fig:mapy}b we plot the coincidences in terms of sum
of wavevector variables. Gaussian fit yields values of $\sigma_x=4.45\pm0.02\ \mathrm{mm}^{-1}$
and $\sigma_y=4.76\pm0.02\ \mathrm{mm}^{-1}$ for $x$ and $y$ dimension,
respectively. Consequently, we can consider the generated entangled
state \cite{Lee2016,Dabrowski2017} to be nearly symmetrical in terms
of $x$ and $y$ spatial dimensions. This wavevector spread is very
close to the limit $2/w$ given by the diffraction at the atomic ensemble
and confirms the quality of the imaging system for conjugate modes.

\textbf{Capacity estimation.} From the fundamental point of view multimode states of light can be
considered either as continuous-variable systems \cite{Tasca2011}
or highly-dimensional entangled states \cite{Fickler2016} offering
large dimensionality of available Hilbert space and in turn providing
high informational capacity. Estimation of the informational capacity
of continuous-variable entangled states of light has attracted some
attention of its own due to broad applications of such states \cite{Moreau2014,Dabrowski2017}. 
Various measures of this capacity have been discussed e.g. on the
information-theoretical basis \cite{Schneeloch2013}. Here we estimate
the number of independent mode pairs observed in S and AS arms using
the Schmidt mode decomposition \cite{Grobe1994,Law2004}. For a single-dimensional
photon pair amplitude given by equation~(\ref{eq:biampl}) and cropped to a finite region we find a
decomposition into the Schmidt modes as:
\begin{equation}
\Psi_y(k_{y,\mathrm{S}},k_{y,\mathrm{AS}})=\sum_{j=0}^{\infty}\lambda_{j}u_{j}(k_{y,\mathrm{S}})v_{j}^{*}(k_{y,\mathrm{AS}}),\label{eq:eigd}
\end{equation}
where $\lambda_{j}$ are singular values corresponding to contributions
of each mode while $u_{j}$ and $v_{j}^{*}$ are orthogonal sets of
eigenfunctions. Effective number of independent mode pairs is given
by $M=1/\sum_{j=0}^{\infty}\lambda_{j}^{4}$ which may be also described
in terms of modes given in an orbital-angular-momentum (OAM) \cite{Ding2013,Nicolas2014,Fickler2016} or
another orthogonal basis.
We find the effective number of modes $M_{x,y}$
is well approximated by an inverse relation $M_{x,y}=0.565\ \kappa/2\sigma_{x,y}$
(see Methods for justification) and obtain $M_{x}=26.7\pm0.1$ and
$M_{y}=24.9\pm0.1$. Finally, for the total number of modes $M$, which
is the product of the number of modes in each spatial dimension, we
get $M=M_{x}M_{y}=665\pm4$. 

\begin{figure*}
\includegraphics[scale=0.78]{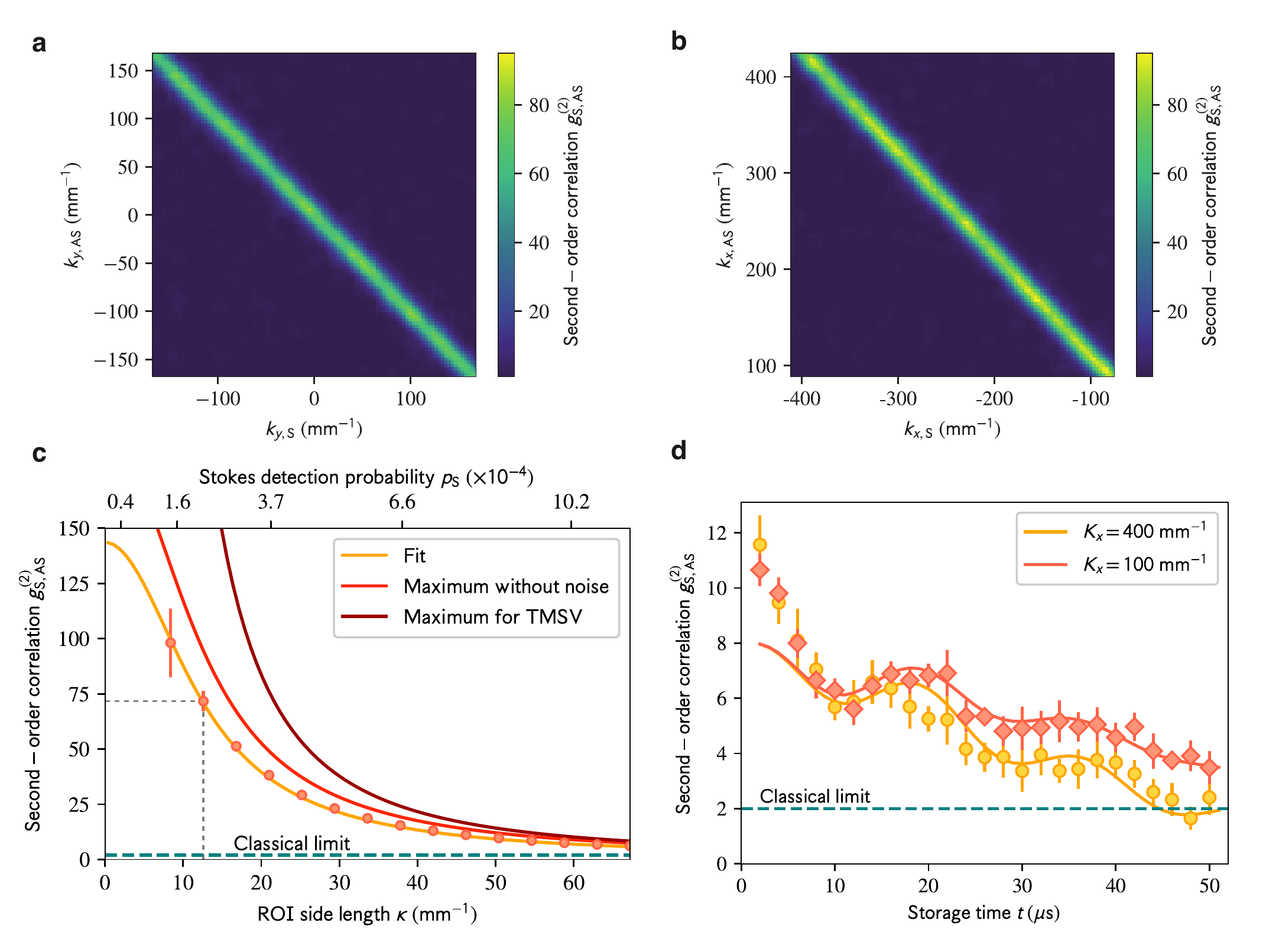}

\caption{\textbf{Nonclassical correlations of photons emitted from the quantum
memory.} \textbf{(a,b)} Second-order cross-correlation function $g_{\mathrm{S,AS}}^{(2)}$
measured for different positions of ROI in S and AS arms, for zero memory storage time. Nonclassical
correlations are observed only between conjugate modes, confirming the
highly-multimode character of our quantum memory. Data corresponds
to S photon probability $p_{\mathrm{{S}}}=2\times10^{-4}$ per ROI. Standard deviation error maps are included as Supplementary~Fig.~1.  \textbf{(c)} $g_{\mathrm{S,AS}}^{(2)}$ for Stokes and anti-Stokes
photons measured at $\tau=0$ storage time using different sizes of ROI in the analysis. Smaller
ROIs correspond to lower $p_{\mathrm{{S}}}$ and consequently give
higher values of $g_{\mathrm{S,AS}}^{(2)}$. Our theoretical prediction for $g_{\mathrm{S,AS}}^{(2)}$ calculated for the measured mode size closely adheres to experimental results (see Methods
for details). Other curves correspond to the maximum value of $g_{\mathrm{S,AS}}^{(2)}$
without noise in the AS arm and the maximum theoretical result for two-mode squeezed vacuum state (TMSV) with given probability $p_{\mathrm{{S}}}$. Grey dashed lines mark the
regime of operation used in the measurement shown in (a,b). \textbf{(d)}
Second-order correlation as a function of storage time, measured for
two different angles of scattering corresponding to stored spin-waves
with different $K_{x}$. Data was taken with a higher than in (a-c) S photon detection
probability of $p_{\mathrm{S}}=1.9\times10^{-3}$ and
thus the value of the correlation function is smaller. Nonclassical correlations
for spin-waves with smaller wavenumber are confirmed for the storage
time $t$ up to $50\ \mu\mathrm{s}$.
Theoretical model of the time-evolution of $g_{\mathrm{S,AS}}^{(2)}$ (solid lines, see Methods for derivation) exhibits good agreement with experimental data,
except for the initial drop that we attribute to an increase of noise fluorescence of thermal atoms. Errobars in panels (c) and (d) correspond to one standard deviation drawn from an ensemble of multiple conjugate region pairs. \textbf{\label{fig:g2}}}
\end{figure*}

\textbf{Nonclassical photon-number correlations.} The presented spatial
correlations at a single-photon level require further analysis to
confirm actual generation of multi-photon quantum states of light.
Quantumness of the correlations (and hence the memory) may be assessed
by looking at the second-order correlation function \cite{Paul1982}:
\begin{equation}
g_{\mathrm{S,AS}}^{(2)}=\frac{{p_{\mathrm{S,AS}}}}{p_{\mathrm{S}}p_{\mathrm{AS}}},
\end{equation}
where $p_{\mathrm{S}}$ and $p_{\mathrm{AS}}$ are the probabilities
of registering a Stokes and an anti-Stokes photon in their respective
regions, while $p_{\mathrm{S,AS}}$ is the Stokes\textendash anti-Stokes
coincidence probability. Since the single mode statistics of Stokes
and anti-Stokes light are thermal \cite{Lee2011}, the maximum value of local $g_{\mathrm{S,S}}^{(2)}$
and $g_{\mathrm{AS,AS}}^{(2)}$ autocorrelation functions is $2$.
Consequently, a value of $g_{\mathrm{S,AS}}^{(2)}>2$ yields violation
of the Cauchy-Schwarz inequality\cite{Paul1982}:
\begin{equation}
R=\frac{[g^{(2)}_\mathrm{S,AS}]^2}{g^{(2)}_\mathrm{S,S}g^{(2)}_\mathrm{AS,AS}}\leq1
\label{eq:CS}
\end{equation}
 and thus proves nonclassical character
of the generated state of light.

To perform the measurements we utilize the photon-number resolving
capability of the I-sCMOS detector \cite{Chrapkiewicz2014}. We verify non-classical photon
number correlations in many modes by selecting a set of ROIs in both
S and AS arms and calculating $g_{\mathrm{S,AS}}^{(2)}$ for all accessible
combinations. Results presented in Fig. \ref{fig:g2}a and b clearly
confirm the multimode capacity discussed in previous sections. For the experimental data presented in Fig. \ref{fig:g2}b we obtain $g_{\mathrm{S,AS}}^{(2)}=72\pm5$ at the diagonal compared with  $g_{\mathrm{S,AS}}^{(2)}=1.0\pm0.4$ for a set of uncorrelated regions, where the errors correspond to one standard deviation. Next, we select
a single pair of square-shaped conjugate ROIs in S and AS arms. Figure
\ref{fig:g2}c presents the measured $g_{\mathrm{S,AS}}^{(2)}$ at $\tau=0$ storage time for
varying size of ROI with a constant photon flux per pixel. With the
decreasing size of ROI the S photon detection probability $p_{\mathrm{{S}}}$
decreases and we observe $g_{\mathrm{S,AS}}^{(2)}$ cross-correlation well above the classical limit of 2 which perfectly matches our theoretical
predictions (see Methods for detailed theory of $g_{\mathrm{S,AS}}^{(2)}$
measured with a multi-pixel detector). We compare
this result with a maximum value achievable without noise in the AS
arm as well as maximum theoretical value for two-mode squeezed vacuum state (TMSV) for the given $p_{\mathrm{S}}$,
achievable only if coherent spatial filtering (using e.g. single-mode
fibers or cavities) is applied.

Even though we expect the photon statistics in S and AS arms to exhibit maximum values of autocorrelation functions of 2, to implicitly demonstrate violation of Cauchy-Schwartz inequality (\ref{eq:CS}), we have performed additional measurements of $g^{(2)}_\mathrm{S,S}$ and $g^{(2)}_\mathrm{S,S}$ using a slightly modified experimental setup (see Methods for details). Due to inherently low number of S-S and AS-AS coincidences we have increased the mean photon number in the S arm to 1.2 obtaining an average value of $R=4.0\pm0.2$, significantly violating inequality (\ref{eq:CS}) and proving both $g^{(2)}_\mathrm{S,S},\:g^{(2)}_\mathrm{AS,AS}\leq2$ (see Supplementary Fig. 2 for spatially resolved maps).

\textbf{Storage capabilities.} Cold atomic ensemble prepared in MOT
typically offers $\mu$s up to ms coherence times, limited mainly by
atomic motion, atom losses and stray magnetic fields. We characterize
the memory storage time by analyzing the $g_{\mathrm{S,AS}}^{(2)}$
correlation function when the read laser is applied after a variable
storage time $t$ following the write pulse. Figure \ref{fig:g2}d
presents the average $g_{\mathrm{S,AS}}^{(2)}$ calculated for 1,000
pairs of correlated square-shaped ROIs with side length $\kappa=21\ \mathrm{mm}^{-1}$
and $p_{\mathrm{{S}}}=1.9\times10^{-3}$ per entire ROI, each comprising approximately 5 modes. Data sets in Fig.
\ref{fig:g2}d correspond to two different angles at which the photons
were scattered, hence spin-waves with different wavevectors \textemdash{}
higher scattering angles (and thus spin-waves with larger wavenumbers)
correspond to shorter decay times. We observe a quantum-beating
oscillation on a double exponential decay of correlations due to the presence
of two types of spin-waves arising as a result of imperfect optical
pumping (see Methods for details). Due to the axial magnetic field of $36\ \mathrm{mG}$, the two types
of spin-waves accumulate different phases over the storage time
which leads to their constructive or destructive interference
at the readout stage. We observe this interference effect as an oscillation
with a Larmor period of $T=2\pi/\omega=19.5\ \mu\mathrm{s}$. For all spin-waves we measure lifetimes larger than $50\ \mu s$. In particular for the clock-transition spin-wave (between $F=1,\ m_{F}=1$ and $F=2,\ m_{F}=-1$ states) with small $K_{x}=100\ \mathrm{mm}^{-1}$ we obtain the lifetime of over $100\ \mu s$.
The main source of decoherence is the random atomic motion governed by the Maxwell-Boltzmann
velocity distribution \cite{Zhao2009}. The sharp drop in $g_{\mathrm{S,AS}}^{(2)}$ in the very beginning
(two initial experimental point) is attributed to increase of noise fluorescence as a result
of an influx of unpumped thermal atoms into the interaction region.  This noise might be eliminated by optical pumping of
thermal atoms or by using a two-stage MOT with differential pumping. See Supplementary Fig. 3 for the measured temporal evolution of noise fluorescence.

\section*{Discussion}
We have demonstrated a quantum memory-enabled source of spatially-structured nonclassical light based on a
principle of wavevector multiplexing. Simultaneous
operation on many collective atomic excitations allows us to generate
a multimode quantum state of light. The memory preserves nonclassical
correlations up to 50 $\mu$s and exhibits excellent noise properties,
in contrast to the hitherto used warm-atomic vapour schemes \cite{Chrapkiewicz2017,Dabrowski2017}.
Simultaneous detection using a state-of-the-art single-photon resolving
camera is an ideal scheme to implement the enhanced photon generation
protocols \cite{Nunn2013,Mazelanik2016,Chrapkiewicz2017}.  Additionally, a two-dimensional detector is both necessary and well-suited to the access high quantum information capacity of multimode states of light, which is unachievable with single-mode fibers \cite{Edgar2012}. 
Furthermore, simultaneous detection of the entire transverse field of view is essential in fundamental tests such as demonstration of the Einstein-Podolsky-Rosen paradox\cite{Einstein1935} without the Bell sampling loophole \cite{Moreau2014}. 

Our results
clearly demonstrate the ability of multimode quantum memory to emit
a single photon with high probability. In particular, we measured S photon detection probability
of $0.21$ and simultaneously extremely low probability of registering
a photon per mode equal $3.8\times10^{-4}$ that drastically minimizes
the probability of generating a photon pair in a single mode and proves
memory efficacy in enhanced generation of photons. Excellent quality
of single photons has been verified through measurements of $g_{\mathrm{S,AS}}^{(2)}$
cross-correlation function. Our quantum memory also exhibits an excellent time-bandwidth
product of more than 500, which is an important figure of merit in terms of probability of retrieving all the photons stored in the memory (see Fig. \ref{fig:multiplex}), as well as prospective integration with time-bin multiplexing \cite{Nunn2013}. We envisage that
hundreds of $\mu$s memory lifetime, contrasted with noise-free yet
low storage-time solutions \cite{Kaczmarek}, and 100 ns operation
time are excellent parameters when it comes to integration with fast
electronic or photonic circuits for real-time feedback \cite{Ma2011,Hall2011}. With these technical difficulties overcame, we expect that the proposed enhanced multi-photon generation protocol would be readily realizable. Integration of existing schemes\cite{Cho2016} with readout efficiency $\chi_\mathrm{R}$ of nearly 0.9 and the probability to generate $n$ photons equal $(\chi_\mathrm{R})^n$, will make our protocol highly competitive. Keeping a low probability of generating a photon--atomic excitation pair per mode $p\approx0.01$ our setup can emit $n=p M \approx 6.6$ photons on average and thus could efficiently generate even 6-photon states in the memory. Consequently, the number of modes $M$ places a fundamental yet possibly distant limit on generating multi-photon states.

The number of available modes is limited by the imaging system. In
a cold atomic ensemble generated within a released MOT we expect that the
final limit for the number of modes will be set by the lifetime of long-wavevector spin-waves
as well as the phase-matching at the retrieval stage. To keep the lifetime
within tens-of-microseconds regime the maximum scattering angles should be smaller than 6 degrees while the phase-matching
happens to place a similar limitation \cite{Surmacz2008,Leszczynski2017}.
We thus predict that the number of readily available modes may reach
thousands under realistic experimental conditions. However, with
novel spin-wave manipulation techniques \cite{Hetet2015} or by placing
the atoms in an optical lattice \cite{Radnaev2010} at least another
order-of-magnitude improvement could be achieved, allowing our setup to serve as a universal platform for quantum state preparation.

While we have focused on the application of our quantum memory as a light
source in multiplexed communication and computation protocols \cite{Nunn2013,Chrapkiewicz2017},
our scheme is also perfectly matched to expedite quantum communication in free-space\cite{Wang2012} or with multimode or multi-core fibers\cite{Richardson2013}, quantum imaging and image processing at the single-photon level, as well as quantum enhanced metrology \cite{Wolfgramm2012,Matthews2016}. Spatial photon-number
quantum correlations are readily applicable in quantum imaging techniques
and the memory capability could help quantum ghost imaging or sub-shot
noise imaging along the way to practical applications \cite{Brida2010,Genovese2016}.
Furthermore, the quantum-beat signal between two spin-wave excitations
demonstrates the ability of our quantum memory to store a superposition
of a few spin-waves in many modes and paves the way towards 
manipulations within and between the Zeeman sublevels as well as with
the spatial degree of freedom.

\section*{Methods}

\textbf{Experimental sequence.} Our experimental sequence depicted
in Fig. \ref{fig:setup}c starts with trapping the atoms in MOT for
$1.4\ \mathrm{ms}$ in an octagonal double-side AR coated glass chamber
(Precision Glassblowing) with cooling laser red-detuned from $5^{2}S_{1/2},\ F=2\rightarrow5^{2}P_{3/2},\ F=3$ transition
by 16 MHz followed by $300\ \mu\mathrm{s}$ phase of polarization
gradient cooling (PGC, cooling laser detuning increased
to 35 MHz by tuning the double-pass acousto-optic modulator) that
brings the temperature from roughly $100\ \mu\mathrm{K}$ to $22\pm2\ \mu\mathrm{K}$.
Fast MOSFET-transistor-based coil switch turns off the coil current
in less than $5\ \mu\mathrm{s}$ from 125 A down to zero, and thus the MOT is turned off during memory operation. Small stray
magnetic fields due to eddy currents take another $200\ \mu\mathrm{s}$
to decay but are compensated to nearly zero in the position of the atomic
ensemble by a shorted compensation coils, which we verified by taking
the free-induction decay measurements \cite{Behbood2013}. Compensation coils maintain a constant axial magnetic field of 36 mG during both the optical pumping and the memory operation stages. The cooling
and trapping period is followed by $40\ \mu\mathrm{{s}}$ stage of
optical pumping carried out by three lasers to ensure emptying of
the memory state $|h\rangle$ and maximizing of Zeeman population in the
$F=1,\ m_{F}=1$ state. Two lasers empty the $F=2$ manifold: a strong
pulse of the read laser (10 mW, depletion stage) as well as another
laser (hyperfine pump \textendash{} HF) tuned to $5^{2}S_{1/2},\ F=2\rightarrow5^{2}P_{1/2},\ F'=2$
transition, incident from four directions (with various polarizations)
with a total power of $10\ \mathrm{mW}$. Another laser (Zeeman pump)
with a power of $7\ \mathrm{mW}$ is resonant to $5^{2}S_{1/2},\ F=1\rightarrow5^{2}P_{3/2},\ F'=1$
transition and transfers the population from $m_{F}=-1$ and $m_{F}=0$
to $m_{F}=1$ state of the $F=1$ manifold.
Dark period of $1\ \mu\mathrm{s}$ follows the optical pumping to ensure all light from lasers
and from the atomic ensemble is extinguished. Next, a $100\ \mathrm{ns}$
write pulse is applied (left-circular polarization, red-detuned by
20 MHz from $5^{2}S_{1/2},\ F=1\rightarrow5^{2}P_{3/2},\ F'=2$ transition). Due to small detunings from respective energy levels the influence of deleterious processes of readout (write-in) with the write (read) laser is negligible.
After a variable memory storage time a $200\ \mathrm{ns}$ read pulse
(right-circular polarization, resonant with $5^{2}S_{1/2},\ F=2\rightarrow5^{2}P_{1/2},\ F=2$
transition) is applied. All lasers are locked to either cooler or
repumper laser through a beat-note offset lock \cite{Lipka2016}.

\textbf{Imaging.} The angles at which the photons are emitted from
the atomic ensemble are imaged on the photocathode of the image intensifier
with two separate (for S and AS) complex telescopes composed of 6
lenses, each having an effective focal length of $f_{\mathrm{eff}}=50\ \mathrm{mm}$.
Total length of the single system is 2 m as optically pumped atomic
filters need to fit along the photons path. The linear size of one pixel of the sCMOS camera corresponds to transverse wavevector size of $2.1\ \mathrm{mm}^{-1}$ or angle of 265 $\mu$rad. Lens apertures limit our
field of view to a ROI with $\kappa=420\ \mathrm{mm}^{-1}$
or total solid angle of $\Omega=2.76\ \mathrm{msr}$, which later
determines maximum number of observable modes. The imaging system was
calibrated with custom Ronchi rulings. 
The I-sCMOS
device, composed of image intensifier (Hamamatsu V7090D) and an sCMOS
sensor (Andor Zyla 5.5 Megapixel), is gated (Photek GM300-3
gating module) only during write-in and readout of atomic excitations.
The sequence is repeated at the rate of
500~Hz which is limited by the frame rate of the sCMOS camera. The combined image intensifier gate duration is 400 ns for both write-in and readout stages. The probability of registering a dark count for the combined S and AS fields of view is approx. $5\times10^{-3}$ \textemdash{} much less than typical photon detection probability of 0.2\textendash{}0.3.

\begin{figure}
\includegraphics[scale=0.8]{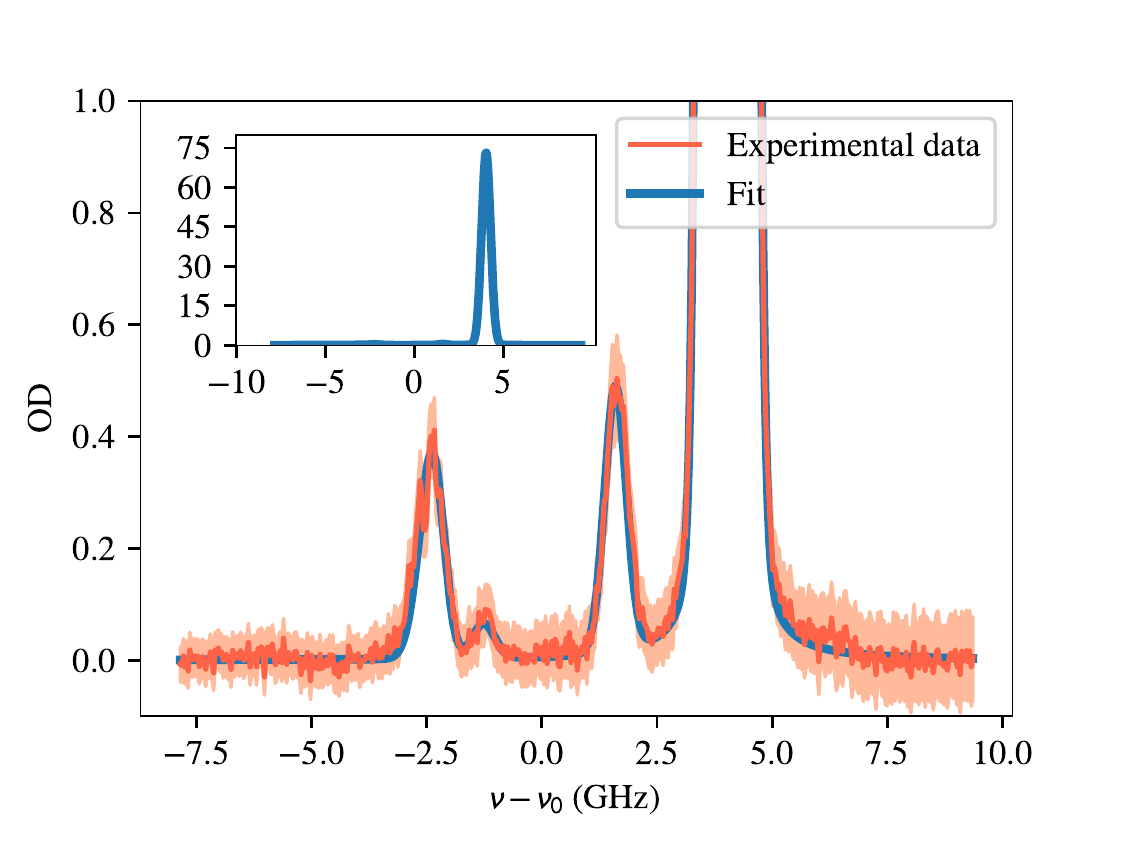}\caption{\textbf{Characterization of the atomic filtering system for Stokes photons. }Data points
correspond to a measurement of absorption of the weak probe beam while fit is the theoretical
prediction of OD based on a Voigt-profile absorption model including
optical pumping. Inset shows the OD in a larger scale (only the fitted
function), demonstrating very high attenuation of write laser light
($\nu-\nu_{0}=4.3$ GHz detuning). Simultaneously, high transmission
of S photons ($\nu-\nu_{0}=-2.5$ GHz detuning) is achieved. Similar
characteristics were also obtained for the AS filter. Detuning is given
with respect to the line centroid at $\nu_0$. Two more central absorption
peaks corresponds to a residual amount of $^{85}$Rb in the filtering
glass cell. Experimental errorbars correspond to one standard mean error derived from many collected spectra. \label{fig:odf}}
\end{figure}

\textbf{Filtering.} Two separate rubidium vapor cells are used to
filter out stray write and read laser light from S and AS photons.
The 10-cm-long cells are paraffin-coated and contain 99.4\% isotopically
pure $^{87}$Rb as well as buffer gases (Precission Glassblowing,
1 torr Kr for both S and AS filters) that keep the pumped atoms in
the interaction region. The cells are pumped with 50 mW of resonant
laser light (with $5^{2}S_{1/2},\ F=2\rightarrow5^{2}P_{1/2}$ and $5^{2}S_{1/2},\ F=1\rightarrow5^{2}P_{3/2}$ for S and AS filters, respectively) in a double-pass configuration with collimated 1-cm-wide
beams. The optical pumping is active at all times except when the
image intensifier gate is open. Importantly, write and read lasers are filtered with
Fabry-P\'erot cavities before illuminating the ensemble to eliminate
amplified spontaneous emission from laser diodes that would not be
filtered with the hot atomic cell. Figure \ref{fig:odf} presents
a characterization of the S filter (we measured the comparable characteristics
also for the AS filter). Both filters are characterized by $\mathrm{OD}>70$
for the laser light and approx. 65\% transmission for single photons
generated inside the atomic quantum memory.

\textbf{Autocorrelation measurement.}
Part of the experimental setup has been modified to allow measurement of autocorrelation functions $g^{(2)}_\mathrm{S,S}$ and $g^{(2)}_\mathrm{AS,AS}$. A high extinction-ratio Wollaston prism was placed in front of the image intensifier and a pair of half-wave plates was used to rotate the polarization of S and AS photons. The Wollaston prism split the photons into two beams (both for S and AS arm) at the 50:50 ratio in the vertical direction, so four distinct regions were observed on the camera (S1, S2, AS1 and AS2). After compensating for the change in angle-of-incidence due refraction at the Wollaston prism we have analyzed the correlations between regions S1-S2 and AS1-AS2 to obtain estimates of auto-correlation functions. The results are presented in Supplementary Fig. 2.

\textbf{Coincident counts.} Let us consider a collection of $M$ squeezed modes pairs. Assuming the probability $p$ of generating
S photon in a single mode and efficiencies in the S and AS arms equal
$\eta_{\mathrm{S}}$ and $\eta_{\mathrm{AS}}\chi_{\mathrm{R}}$, respectively,
with $\chi_{\mathrm{{R}}}$ being the retrieval efficiency, we obtain the probability of registering a
coincidence from any two conjugate modes $pM\eta_{\mathrm{S}}\eta_{\mathrm{AS}}\chi_{\mathrm{R}}$. 
If we now consider a pair of square-shaped ROIs with the side length $\kappa$ containing $M$ modes for which we again assume the probability $p$ per mode to generate a photon
pair, the coincidence rate is reduced, as not all coincidences will fall into the ROI. This effect is more pronounced for the smaller size of ROI.  In particular, if we consider that the S photon is detected inside its respective ROI, we seek the probability that its conjugate AS photon will be detected in conjugate ROI (i.e. with conjugate center). We may calculate this probability by considering photon pairs distributed in momentum space according
to equation (\ref{eq:biampl}). By considering S photons in the given ROI
we calculate the conditional probability $f(\kappa)$ of registering AS photon in the conjugate
ROI in the AS arm, which gives us:
\begin{eqnarray}
\nonumber
f(\kappa)=
\left(\int_{-\frac{\kappa}{2}}^{\frac{\kappa}{2}}\mathrm{d}k_{\mathrm{S}}\int_{-\frac{\kappa}{2}}^{\frac{\kappa}{2}}\mathrm{d}k_{\mathrm{AS}}\frac{{1}}{\sqrt{2\pi}\kappa\sigma}e^{-(k_{\mathrm{S}}+k_{\mathrm{AS}})^{2}/2\sigma^{2}}\right)^{2}\\
=\left(\text{ erf}\left(\frac{\sqrt{2}\kappa}{2\sigma}\right)+\sqrt{\frac{1}{2\pi}}\frac{2\sigma}{\kappa}\left(e^{-\kappa^{2}/2\sigma^{2}}-1\right)\right)^{2},\ \ \ \ \ \ 
\end{eqnarray}
where squaring is due to the two-dimensional character of the problem. Finally, to estimate the net coincidence probability we additionally consider the total number of accidental coincidences which is very well approximated by a product of probabilities in S and AS arms $p_\mathrm{S} p_\mathrm{AS}$ \cite{Chen2006,Zhao2009,Albrecht2015,Laplane2017,Dabrowski2017}.

The net S\textendash AS coincidence probability thus equals:
\begin{equation}
p_{\mathrm{{S,AS}}}=pMf(\kappa)\eta_{\mathrm{S}}\eta_{\mathrm{AS}}\chi_{\mathrm{R}}+p_\mathrm{S} p_\mathrm{AS}.
\end{equation}

\textbf{Second order correlation.} We model the evolution of $g_{\mathrm{S,AS}}^{(2)}$
correlation function following \emph{Zhao et al.} \cite{Zhao2009}, but
including the effect of interference of different spin-waves as well
as the reduced number of coincidence counts due to incoherent spatial filtering.
Finally, we end up with the following expression for the second-order
correlation function:
\begin{equation}
g_{S,AS}^{(2)}=1+\frac{{p\eta_\mathrm{S}\eta_\mathrm{AS}\chi_{\mathrm{{R}}}(t)f(\kappa)}}{p\eta_\mathrm{S}(p\eta_\mathrm{AS}\chi_{\mathrm{{R}}}(t)+\xi)},
\end{equation}
where $\xi$ is a contribution of noise in the AS arm. The retrieval efficiency
is modeled as an interference of two fields arising due to two atomic
coherences by the following time-dependent expression:
\begin{equation}
\chi_{\mathrm{R}}(t)=|\alpha_{1}\exp(-t^{2}/2\tau_{1}^{2})+\alpha_{2}\exp(i\omega t)\exp(-t^{2}/2\tau_{2}^{2})|^{2},
\end{equation}
where $\alpha_{1}$ and $\alpha_{2}$ are contributions of spin-waves
between $F=1,\ m_{F}=1\leftrightarrow F=2,\ m_{F}=-1$ and $F=1,\ m_{F}=0\leftrightarrow F=2,\ m_{F}=-2$ transitions,
respectively. The fit yields $\alpha_{1}=0.58$ and $\alpha_{2}=0.04$,
clearly confirming dominant role of the clock-transition spin-wave.
The relative phase between the two spin-waves changes as one of them
accumulates additional phase due to a Zeeman energy shift $\hbar\omega=2\pi\hbar\times51\ \mathrm{kHz}$
in the axial magnetic field of 36 mG. 
The lifetimes
are bounded by wavevector-dependent decoherence rate $\Gamma_{D}=|\mathbf{{K}}|v$,
with $v=\sqrt{\frac{{k_{\mathrm{B}}T}}{m_{\mathrm{Rb}}}}\approx1.45\ \mathrm{{cm\ s^{-1}}}$.

\begin{figure}
\begin{centering}
\includegraphics[scale=0.7]{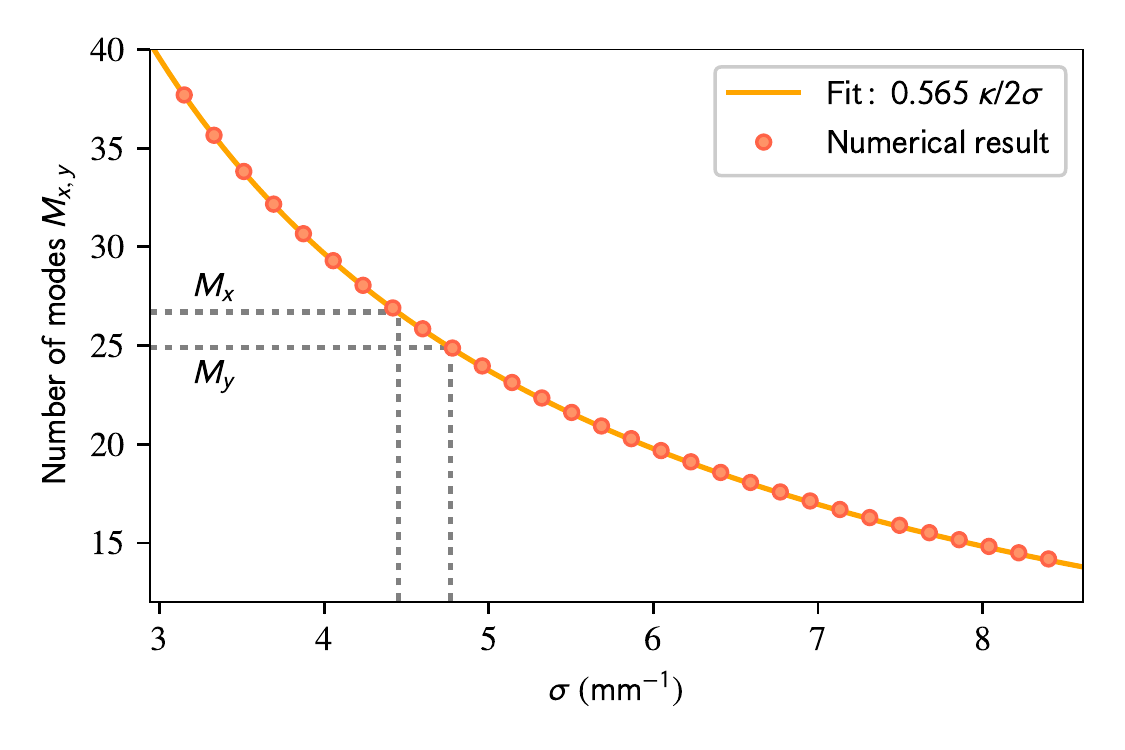}
\par\end{centering}
\centering{}\caption{\textbf{Result of the eigenmode decomposition for the number of modes.
}Dots represent the results from numerical decomposition while the
solid line is the simplified prediction of $0.565\:\kappa/2\sigma$.
Dotted grey lines correspond to values of $\sigma_{x,y}$ we obtain
in our experimental setup and corresponding numbers of modes $M_{x}$
and $M_{y}$.\label{fig:K-sigma}}
\end{figure}

\textbf{Uncertainty estimation.}
For the data presented in Fig. 4a to obtain a single $g^{(2)}_\mathrm{S,AS}$ map we have selected 100 regions in a column (oriented in the $y$-direction) in S and AS arms and calculated value of $g^{(2)}_\mathrm{S,AS}$ for each pair of regions (note these regions are partially overlapping). We collect these results for 25 different conjugate positions of columns in the $x$-direction (i.e. $k_{x,\mathrm{S}}+k_{x,\mathrm{AS}}=0$) and estimate mean and standard deviation. A completely analogous procedure was applied to obtain Fig. 4b and corresponding errorbars. Maps of standard deviation are included as Supplementary Fig. 1.
To calculate the value of $g^{(2)}_\mathrm{S,AS}$ for Fig. 4c we have additionally averaged over all conjugate regions (corresponding to averaging over the diagonal in Fig. 4b) and inferring the errorbars (one standard deviation).
Similar procedure was used to calculate values and standard deviations in Fig. 4d, however we used far less regions to average as we required that regions correspond to appropriate spin-wave wavevector $\mathbf{K}_x$. Furthermore, in this measurement less frames were collected for each point and thus we obtain relatively high uncertainty. 

\textbf{Eigenmode decomposition.} To correctly determine the number
of modes we use a similar procedure as proposed by \emph{Law} and \emph{Eberly} \cite{Law2004}.
Focusing on one dimension, we generate a normalized biphoton amplitudes
according to equation (\ref{eq:biampl}), with various widths $\sigma$, on
a square two-dimensional $k_{y,\mathrm{{S}}}$\textendash $k_{y,\mathrm{{AS}}}$
 grid. We numerically find the eigenmode decomposition of the generated
matrix and calculate the number of modes according to equation (\ref{eq:eigd}).
Figure \ref{fig:K-sigma} presents example of results for $\kappa=420\ \mathrm{{mm}}^{-1}$
while the solid line corresponds to a fit of $A\kappa/2\sigma$ relation
which we verified numerically for various sets of parameters and obtained $A=0.565$.
Note that for a biphoton amplitude on a rectangular (non-square) grid
numerical singular value decomposition (SVD) might be used to give
similar results.


\begin{thebibliography}{10}
\expandafter\ifx\csname url\endcsname\relax
  \def\url#1{\texttt{#1}}\fi
\expandafter\ifx\csname urlprefix\endcsname\relax\def\urlprefix{URL }\fi
\providecommand{\bibinfo}[2]{#2}
\providecommand{\eprint}[2][]{\url{#2}}
 
\bibitem{Wang2012}
\bibinfo{author}{Wang, J.} \emph{et~al.}
\newblock \bibinfo{title}{{Terabit free-space data transmission employing
  orbital angular momentum multiplexing}}.
\newblock \emph{\bibinfo{journal}{Nature Photonics}}
  \textbf{\bibinfo{volume}{6}}, \bibinfo{pages}{488--496}
  (\bibinfo{year}{2012}).

\bibitem{Richardson2013}
\bibinfo{author}{Richardson, D.~J.}, \bibinfo{author}{Fini, J.~M.} \&
  \bibinfo{author}{Nelson, L.~E.}
\newblock \bibinfo{title}{{Space-division multiplexing in optical fibres}}.
\newblock \emph{\bibinfo{journal}{Nature Photonics}}
  \textbf{\bibinfo{volume}{7}}, \bibinfo{pages}{354--362}
  (\bibinfo{year}{2013}).

\bibitem{Munro2010}
\bibinfo{author}{Munro, W.~J.}, \bibinfo{author}{Harrison, K.~A.},
  \bibinfo{author}{Stephens, A.~M.}, \bibinfo{author}{Devitt, S.~J.} \&
  \bibinfo{author}{Nemoto, K.}
\newblock \bibinfo{title}{{From quantum fusiliers to high-performance
  networks}}.
\newblock \emph{\bibinfo{journal}{Nature Photonics}}
  \textbf{\bibinfo{volume}{4}}, \bibinfo{pages}{792--796}
  (\bibinfo{year}{2010}).

\bibitem{Surmacz2008}
\bibinfo{author}{Surmacz, K.} \emph{et~al.}
\newblock \bibinfo{title}{{Efficient spatially resolved multimode quantum
  memory}}.
\newblock \emph{\bibinfo{journal}{Physical Review A}}
  \textbf{\bibinfo{volume}{78}}, \bibinfo{pages}{033806}
  (\bibinfo{year}{2008}).

\bibitem{Lan2009}
\bibinfo{author}{Lan, S.-Y.} \emph{et~al.}
\newblock \bibinfo{title}{{A multiplexed quantum memory}}.
\newblock \emph{\bibinfo{journal}{Optics Express}}
  \textbf{\bibinfo{volume}{17}}, \bibinfo{pages}{13639--13645}
  (\bibinfo{year}{2009}).

\bibitem{Dai2012a}
\bibinfo{author}{Dai, H.-N.} \emph{et~al.}
\newblock \bibinfo{title}{{Holographic Storage of Biphoton Entanglement}}.
\newblock \emph{\bibinfo{journal}{Physical Review Letters}}
  \textbf{\bibinfo{volume}{108}}, \bibinfo{pages}{210501}
  (\bibinfo{year}{2012}).
  
\bibitem{Ding2013}
\bibinfo{author}{Ding, D.~-S.}, \bibinfo{author}{Zhou, Z.~-Y.},
  \bibinfo{author}{Shi, B.~-S.} \&
  \bibinfo{author}{Guo, G.~-C.}
\newblock \bibinfo{title}{{Single-photon-level quantum image memory based on cold atomic ensembles}}.
\newblock \emph{\bibinfo{journal}{Nature Communications}}
  \textbf{\bibinfo{volume}{4}}, \bibinfo{pages}{252}
  (\bibinfo{year}{2013}).

\bibitem{Nicolas2014}
\bibinfo{author}{Nicolas, A.} \emph{et~al.}
\newblock \bibinfo{title}{{A quantum memory for orbital angular momentum
  photonic qubits}}.
\newblock \emph{\bibinfo{journal}{Nature Photonics}}
  \textbf{\bibinfo{volume}{8}}, \bibinfo{pages}{234--238}
  (\bibinfo{year}{2013}).

\bibitem{Parigi2015}
\bibinfo{author}{Parigi, V.} \emph{et~al.}
\newblock \bibinfo{title}{{Storage and retrieval of vector beams of light in a
  multiple-degree-of-freedom quantum memory}}.
\newblock \emph{\bibinfo{journal}{Nature Communications}}
  \textbf{\bibinfo{volume}{6}}, \bibinfo{pages}{7706} (\bibinfo{year}{2015}).

\bibitem{Lee2016}
\bibinfo{author}{Lee, J.-C.}, \bibinfo{author}{Park, K.-K.},
  \bibinfo{author}{Zhao, T.-M.} \& \bibinfo{author}{Kim, Y.-H.}
\newblock \bibinfo{title}{{Einstein-Podolsky-Rosen Entanglement of Narrow-Band
  Photons from Cold Atoms}}.
\newblock \emph{\bibinfo{journal}{Physical Review Letters}}
  \textbf{\bibinfo{volume}{117}}, \bibinfo{pages}{250501}
  (\bibinfo{year}{2016}).

\bibitem{Chen2016}
\bibinfo{author}{Chen, L.} \emph{et~al.}
\newblock \bibinfo{title}{{Controllably releasing long-lived quantum memory for
  photonic polarization qubit into multiple spatially-separate photonic
  channels}}.
\newblock \emph{\bibinfo{journal}{Scientific Reports}}
  \textbf{\bibinfo{volume}{6}}, \bibinfo{pages}{33959} (\bibinfo{year}{2016}).

\bibitem{Dabrowski2017}
\bibinfo{author}{D{\k{a}}browski, M.}, \bibinfo{author}{Parniak, M.} \&
  \bibinfo{author}{Wasilewski, W.}
\newblock \bibinfo{title}{{Einstein-Podolsky-Rosen paradox in a hybrid
  bipartite system}}.
\newblock \emph{\bibinfo{journal}{Optica}} \textbf{\bibinfo{volume}{4}},
  \bibinfo{pages}{272--275} (\bibinfo{year}{2017}).

\bibitem{Clausen2011}
\bibinfo{author}{Clausen, C.} \emph{et~al.}
\newblock \bibinfo{title}{{Quantum Storage of Photonic Entanglement in a
  Crystal}}.
\newblock \emph{\bibinfo{journal}{Nature}} \textbf{\bibinfo{volume}{469}},
  \bibinfo{pages}{508--511} (\bibinfo{year}{2010}).

\bibitem{Collins2013}
\bibinfo{author}{Collins, M.} \emph{et~al.}
\newblock \bibinfo{title}{{Integrated spatial multiplexing of heralded
  single-photon sources}}.
\newblock \emph{\bibinfo{journal}{Nature Communications}}
  \textbf{\bibinfo{volume}{4}}, \bibinfo{pages}{3413--3415}
  (\bibinfo{year}{2013}).

\bibitem{Humphreys2014}
\bibinfo{author}{Humphreys, P.~C.} \emph{et~al.}
\newblock \bibinfo{title}{{Continuous-variable quantum computing in optical
  time-frequency modes using quantum memories}}.
\newblock \emph{\bibinfo{journal}{Physical Review Letters}}
  \textbf{\bibinfo{volume}{113}}, \bibinfo{pages}{130502}
  (\bibinfo{year}{2014}).

\bibitem{Gundogan2015}
\bibinfo{author}{Gundogan, M.}, \bibinfo{author}{Ledingham, P.~M.},
  \bibinfo{author}{Kutluer, K.}, \bibinfo{author}{Mazzera, M.} \&
  \bibinfo{author}{{De Riedmatten}, H.}
\newblock \bibinfo{title}{{Solid State Spin-Wave Quantum Memory for Time-Bin
  Qubits}}.
\newblock \emph{\bibinfo{journal}{Physical Review Letters}}
  \textbf{\bibinfo{volume}{114}}, \bibinfo{pages}{230501}
  (\bibinfo{year}{2015}).

\bibitem{Cho2016}
\bibinfo{author}{Cho, Y.~W.} \emph{et~al.}
\newblock \bibinfo{title}{{Highly efficient optical quantum memory with long
  coherence time in cold atoms}}.
\newblock \emph{\bibinfo{journal}{Optica}} \textbf{\bibinfo{volume}{3}},
  \bibinfo{pages}{100--107} (\bibinfo{year}{2016}).

\bibitem{Xiong2016}
\bibinfo{author}{Xiong, C.} \emph{et~al.}
\newblock \bibinfo{title}{{Active temporal multiplexing of indistinguishable
  heralded single photons}}.
\newblock \emph{\bibinfo{journal}{Nature Communications}}
  \textbf{\bibinfo{volume}{7}}, \bibinfo{pages}{10853} (\bibinfo{year}{2016}).
  
\bibitem{Sinclair2014}
\bibinfo{author}{Sinclair, N.} \emph{et~al.}
\newblock \bibinfo{title}{{Spectral multiplexing for scalable quantum photonics using an atomic frequency comb quantum memory and feed-forward control}}.
\newblock \emph{\bibinfo{journal}{Physical Review Letters}}
  \textbf{\bibinfo{volume}{113}}, \bibinfo{pages}{053603}
  (\bibinfo{year}{2014}).
  
\bibitem{Puigibert2017}
\bibinfo{author}{Grimau Puigibert, M.} \emph{et~al.}
\newblock \bibinfo{title}{{Heralded Single Photons Based on Spectral Multiplexing and Feed-Forward Control}}.
\newblock \emph{\bibinfo{journal}{Physical Review Letters}}
  \textbf{\bibinfo{volume}{119}}, \bibinfo{pages}{083601}
  (\bibinfo{year}{2017}).

\bibitem{Duan2001}
\bibinfo{author}{Duan, L.-M.}, \bibinfo{author}{Lukin, M.},
  \bibinfo{author}{Cirac, I.} \& \bibinfo{author}{Zoller, P.}
\newblock \bibinfo{title}{{Long-distance quantum communication with atomic
  ensembles and linear optics}}.
\newblock \emph{\bibinfo{journal}{Nature}} \textbf{\bibinfo{volume}{414}},
  \bibinfo{pages}{413--418} (\bibinfo{year}{2001}).

\bibitem{Simon2007}
\bibinfo{author}{Simon, C.} \emph{et~al.}
\newblock \bibinfo{title}{{Quantum repeaters with photon pair sources and
  multimode memories}}.
\newblock \emph{\bibinfo{journal}{Physical Review Letters}}
  \textbf{\bibinfo{volume}{98}}, \bibinfo{pages}{190503}
  (\bibinfo{year}{2007}).
 
\bibitem{Kutluer2017}
\bibinfo{author}{Kutluer, K.}, \bibinfo{author}{Mazzera, M.} \&
  \bibinfo{author}{de~Riedmatten, H.}
\newblock \bibinfo{title}{{Solid-State Source of Nonclassical Photon Pairs with
  Embedded Multimode Quantum Memory}}.
\newblock \emph{\bibinfo{journal}{Physical Review Letters}}
  \textbf{\bibinfo{volume}{118}}, \bibinfo{pages}{210502}
  (\bibinfo{year}{2017}).

\bibitem{Laplane2017}
\bibinfo{author}{Laplane, C.}, \bibinfo{author}{Jobez, P.},
  \bibinfo{author}{Etesse, J.}, \bibinfo{author}{Gisin, N.} \&
  \bibinfo{author}{Afzelius, M.}
\newblock \bibinfo{title}{{Multimode and Long-Lived Quantum Correlations
  Between Photons and Spins in a Crystal}}.
\newblock \emph{\bibinfo{journal}{Physical Review Letters}}
  \textbf{\bibinfo{volume}{118}}, \bibinfo{pages}{210501}
  (\bibinfo{year}{2017}).

\bibitem{Ma2011}
\bibinfo{author}{Ma, X.~S.}, \bibinfo{author}{Zotter, S.},
  \bibinfo{author}{Kofler, J.}, \bibinfo{author}{Jennewein, T.} \&
  \bibinfo{author}{Zeilinger, A.}
\newblock \bibinfo{title}{{Experimental generation of single photons via active
  multiplexing}}.
\newblock \emph{\bibinfo{journal}{Physical Review A}}
  \textbf{\bibinfo{volume}{83}}, \bibinfo{pages}{043814}
  (\bibinfo{year}{2011}).

\bibitem{Nunn2013}
\bibinfo{author}{Nunn, J.} \emph{et~al.}
\newblock \bibinfo{title}{{Enhancing multiphoton rates with quantum memories}}.
\newblock \emph{\bibinfo{journal}{Physical Review Letters}}
  \textbf{\bibinfo{volume}{110}}, \bibinfo{pages}{133601}
  (\bibinfo{year}{2013}).

\bibitem{Chrapkiewicz2017}
\bibinfo{author}{Chrapkiewicz, R.}, \bibinfo{author}{D{\k{a}}browski, M.} \&
  \bibinfo{author}{Wasilewski, W.}
\newblock \bibinfo{title}{{High-Capacity Angularly Multiplexed Holographic
  Memory Operating at the Single-Photon Level}}.
\newblock \emph{\bibinfo{journal}{Physical Review Letters}}
  \textbf{\bibinfo{volume}{118}}, \bibinfo{pages}{063603}
  (\bibinfo{year}{2017}).

\bibitem{Wolfgramm2012}
\bibinfo{author}{Wolfgramm, F.}, \bibinfo{author}{Vitelli, C.},
  \bibinfo{author}{Beduini, F.~A.}, \bibinfo{author}{Godbout, N.} \&
  \bibinfo{author}{Mitchell, M.~W.}
\newblock \bibinfo{title}{{Entanglement-enhanced probing of a delicate material
  system}}.
\newblock \emph{\bibinfo{journal}{Nature Photonics}}
  \textbf{\bibinfo{volume}{7}}, \bibinfo{pages}{28--32} (\bibinfo{year}{2012}).

\bibitem{Matthews2016}
\bibinfo{author}{Matthews, J.~C.} \emph{et~al.}
\newblock \bibinfo{title}{{Towards practical quantum metrology with photon
  counting}}.
\newblock \emph{\bibinfo{journal}{npj Quantum Information}}
  \textbf{\bibinfo{volume}{2}}, \bibinfo{pages}{16023} (\bibinfo{year}{2016}).

\bibitem{Kok2007}
\bibinfo{author}{Kok, P.} \emph{et~al.}
\newblock \bibinfo{title}{{Linear optical quantum computing with photonic
  qubits}}.
\newblock \emph{\bibinfo{journal}{Reviews of Modern Physics}}
  \textbf{\bibinfo{volume}{79}}, \bibinfo{pages}{135--174}
  (\bibinfo{year}{2007}).

\bibitem{Edgar2012}
\bibinfo{author}{Edgar, M.} \emph{et~al.}
\newblock \bibinfo{title}{{Imaging high-dimensional spatial entanglement with a
  camera}}.
\newblock \emph{\bibinfo{journal}{Nature Communications}}
  \textbf{\bibinfo{volume}{3}}, \bibinfo{pages}{984} (\bibinfo{year}{2012}).
  
\bibitem{Moreau2014}
\bibinfo{author}{Moreau, P.~A.}, \bibinfo{author}{Devaux, F.} \&
  \bibinfo{author}{Lantz, E.}
\newblock \bibinfo{title}{{Einstein-Podolsky-Rosen Paradox in Twin Images}}.
\newblock \emph{\bibinfo{journal}{Physical Review Letters}}
  \textbf{\bibinfo{volume}{113}}, \bibinfo{pages}{160401}
  (\bibinfo{year}{2014}).

\bibitem{Krenn2014}
\bibinfo{author}{Krenn, M.} \emph{et~al.}
\newblock \bibinfo{title}{{Generation and confirmation of a (100 x
  100)-dimensional entangled quantum system}}.
\newblock \emph{\bibinfo{journal}{Proceedings of the National Academy of
  Sciences}} \textbf{\bibinfo{volume}{111}}, \bibinfo{pages}{6243--6247}
  (\bibinfo{year}{2014}).

\bibitem{Roslund2013}
\bibinfo{author}{Roslund, J.}, \bibinfo{author}{de~Ara{\'{u}}jo, R.~M.},
  \bibinfo{author}{Jiang, S.}, \bibinfo{author}{Fabre, C.} \&
  \bibinfo{author}{Treps, N.}
\newblock \bibinfo{title}{{Wavelength-multiplexed quantum networks with
  ultrafast frequency combs}}.
\newblock \emph{\bibinfo{journal}{Nature Photonics}}
  \textbf{\bibinfo{volume}{8}}, \bibinfo{pages}{109--112}
  (\bibinfo{year}{2013}).

\bibitem{Kaneda2015}
\bibinfo{author}{Kaneda, F.} \emph{et~al.}
\newblock \bibinfo{title}{{Time-multiplexed heralded single-photon source}}.
\newblock \emph{\bibinfo{journal}{Optica}} \textbf{\bibinfo{volume}{2}},
  \bibinfo{pages}{1010--1013} (\bibinfo{year}{2015}).

\bibitem{Xie2015}
\bibinfo{author}{Xie, Z.} \emph{et~al.}
\newblock \bibinfo{title}{{Harnessing high-dimensional hyperentanglement
  through a biphoton frequency comb}}.
\newblock \emph{\bibinfo{journal}{Nature Photonics}}
  \textbf{\bibinfo{volume}{9}}, \bibinfo{pages}{536--542}
  (\bibinfo{year}{2015}).

\bibitem{Reimer2016}
\bibinfo{author}{Reimer, C.} \emph{et~al.}
\newblock \bibinfo{title}{{Generation of multiphoton entangled quantum states
  by means of integrated frequency combs}}.
\newblock \emph{\bibinfo{journal}{Science}} \textbf{\bibinfo{volume}{351}},
  \bibinfo{pages}{1176--1180} (\bibinfo{year}{2016}).

\bibitem{Cai2017}
\bibinfo{author}{Cai, Y.} \emph{et~al.}
\newblock \bibinfo{title}{{Multimode entanglement in reconfigurable graph
  states using optical frequency combs}}.
\newblock \emph{\bibinfo{journal}{Nature Communications}}
  \textbf{\bibinfo{volume}{8}}, \bibinfo{pages}{15645} (\bibinfo{year}{2017}).

\bibitem{Kaneda2017}
\bibinfo{author}{Kaneda, F.}, \bibinfo{author}{Xu, F.},
  \bibinfo{author}{Chapman, J.} \& \bibinfo{author}{Kwiat, P.~G.}
\newblock \bibinfo{title}{{Quantum-memory-assisted multi-photon generation for
  efficient quantum information processing}}.
\newblock \emph{\bibinfo{journal}{Optica}} \textbf{\bibinfo{volume}{4}},
  \bibinfo{pages}{1034--1037} (\bibinfo{year}{2017}).

\bibitem{Tang2015}
\bibinfo{author}{Tang, J.-S.} \emph{et~al.}
\newblock \bibinfo{title}{{Storage of multiple single-photon pulses emitted
  from a quantum dot in a solid-state quantum memory}}.
\newblock \emph{\bibinfo{journal}{Nature Communications}}
  \textbf{\bibinfo{volume}{6}}, \bibinfo{pages}{8652} (\bibinfo{year}{2015}).

\bibitem{Pu2017}
\bibinfo{author}{Pu, Y.-F.} \emph{et~al.}
\newblock \bibinfo{title}{{Experimental realization of a multiplexed quantum
  memory with 225 individually accessible memory cells}}.
\newblock \emph{\bibinfo{journal}{Nature Communications}}
  \textbf{\bibinfo{volume}{8}}, \bibinfo{pages}{15359} (\bibinfo{year}{2017}).

\bibitem{Boyer2008}
\bibinfo{author}{Boyer, V.}, \bibinfo{author}{Marino, A.~M.},
  \bibinfo{author}{Pooser, R.~C.} \& \bibinfo{author}{Lett, P.~D.}
\newblock \bibinfo{title}{{Entangled Images from Four-Wave Mixing}}.
\newblock \emph{\bibinfo{journal}{Science}} \textbf{\bibinfo{volume}{321}},
  \bibinfo{pages}{544--547} (\bibinfo{year}{2008}).

\bibitem{Brida2010}
\bibinfo{author}{Brida, G.}, \bibinfo{author}{Genovese, M.} \&
  \bibinfo{author}{{Ruo Berchera}, I.}
\newblock \bibinfo{title}{{Experimental realization of sub-shot-noise quantum
  imaging}}.
\newblock \emph{\bibinfo{journal}{Nature Photonics}}
  \textbf{\bibinfo{volume}{4}}, \bibinfo{pages}{227--230}
  (\bibinfo{year}{2010}).
  
\bibitem{Genovese2016}
\bibinfo{author}{Genovese, M.}
\newblock \bibinfo{title}{{Real applications of quantum imaging}}.
\newblock \emph{\bibinfo{journal}{Journal of Optics}}
  \textbf{\bibinfo{volume}{18}}, \bibinfo{pages}{073002}
  (\bibinfo{year}{2016}).

  
\bibitem{Mazelanik2016}
\bibinfo{author}{Mazelanik, M.}, \bibinfo{author}{D{\k{a}}browski, M.} \&
  \bibinfo{author}{Wasilewski, W.}
\newblock \bibinfo{title}{{Correlation steering in the angularly multimode
  Raman atomic memory}}.
\newblock \emph{\bibinfo{journal}{Optics Express}}
  \textbf{\bibinfo{volume}{24}}, \bibinfo{pages}{21995--22003}
  (\bibinfo{year}{2016}).

\bibitem{Hosseini2009}
\bibinfo{author}{Hosseini, M.} \emph{et~al.}
\newblock \bibinfo{title}{{Coherent optical pulse sequencer for quantum
  applications.}}
\newblock \emph{\bibinfo{journal}{Nature}} \textbf{\bibinfo{volume}{461}},
  \bibinfo{pages}{241--5} (\bibinfo{year}{2009}).

\bibitem{Zhang2012}
\bibinfo{author}{Zhang, S.} \emph{et~al.}
\newblock \bibinfo{title}{{A dark-line two-dimensional magneto-optical trap of
  85Rb atoms with high optical depth}}.
\newblock \emph{\bibinfo{journal}{Review of Scientific Instruments}}
  \textbf{\bibinfo{volume}{83}}, \bibinfo{pages}{073102}
  (\bibinfo{year}{2012}).

\bibitem{Chrapkiewicz2014}
\bibinfo{author}{Chrapkiewicz, R.}, \bibinfo{author}{Wasilewski, W.} \&
  \bibinfo{author}{Banaszek, K.}
\newblock \bibinfo{title}{{High-fidelity spatially resolved multiphoton
  counting for quantum imaging applications.}}
\newblock \emph{\bibinfo{journal}{Optics Letters}}
  \textbf{\bibinfo{volume}{39}}, \bibinfo{pages}{5090--5093}
  (\bibinfo{year}{2014}).

\bibitem{Tasca2011}
\bibinfo{author}{Tasca, D.~S.}, \bibinfo{author}{Gomes, R.~M.},
  \bibinfo{author}{Toscano, F.}, \bibinfo{author}{{Souto Ribeiro}, P.~H.} \&
  \bibinfo{author}{Walborn, S.~P.}
\newblock \bibinfo{title}{{Continuous-variable quantum computation with spatial
  degrees of freedom of photons}}.
\newblock \emph{\bibinfo{journal}{Physical Review A}}
  \textbf{\bibinfo{volume}{83}}, \bibinfo{pages}{052325}
  (\bibinfo{year}{2011}).

\bibitem{Fickler2016}
\bibinfo{author}{Fickler, R.}, \bibinfo{author}{Campbell, G.},
  \bibinfo{author}{Buchler, B.}, \bibinfo{author}{Lam, P.~K.} \&
  \bibinfo{author}{Zeilinger, A.}
\newblock \bibinfo{title}{{Quantum entanglement of angular momentum states with
  quantum numbers up to 10,010}}.
\newblock \emph{\bibinfo{journal}{Proceedings of the National Academy of
  Sciences}} \textbf{\bibinfo{volume}{113}}, \bibinfo{pages}{13642--13647}
  (\bibinfo{year}{2016}).

\bibitem{Schneeloch2013}
\bibinfo{author}{Schneeloch, J.}, \bibinfo{author}{Dixon, P.~B.},
  \bibinfo{author}{Howland, G.~A.}, \bibinfo{author}{Broadbent, C.~J.} \&
  \bibinfo{author}{Howell, J.~C.}
\newblock \bibinfo{title}{{Violation of Continuous-Variable
  Einstein-Podolsky-Rosen Steering with Discrete Measurements}}.
\newblock \emph{\bibinfo{journal}{Physical Review Letters}}
  \textbf{\bibinfo{volume}{110}}, \bibinfo{pages}{130407}
  (\bibinfo{year}{2013}).

\bibitem{Grobe1994}
\bibinfo{author}{Grobe, R.}, \bibinfo{author}{Rz{\k{a}}{\.{z}}ewski, K.} \&
  \bibinfo{author}{Eberly, J.~H.}
\newblock \bibinfo{title}{{Measure of electron-electron correlation in atomic
  physics}}.
\newblock \emph{\bibinfo{journal}{Journal of Physics B}}
  \textbf{\bibinfo{volume}{27}}, \bibinfo{pages}{L503--L508}
  (\bibinfo{year}{1999}).

\bibitem{Law2004}
\bibinfo{author}{Law, C.~K.} \& \bibinfo{author}{Eberly, J.~H.}
\newblock \bibinfo{title}{{Analysis and interpretation of high transverse
  entanglement in optical parametric down conversion}}.
\newblock \emph{\bibinfo{journal}{Physical Review Letters}}
  \textbf{\bibinfo{volume}{92}}, \bibinfo{pages}{127903--1}
  (\bibinfo{year}{2004}).

\bibitem{Paul1982}
\bibinfo{author}{Paul, H.}
\newblock \bibinfo{title}{{Photon antibunching}}.
\newblock \emph{\bibinfo{journal}{Reviews of Modern Physics}}
  \textbf{\bibinfo{volume}{54}}, \bibinfo{pages}{1061--1102}
  (\bibinfo{year}{1982}).

\bibitem{Lee2011}
\bibinfo{author}{Lee, K.~C.} \emph{et~al.}
\newblock \bibinfo{title}{{Macroscopic non-classical states and terahertz
  quantum processing in room-temperature diamond}}.
\newblock \emph{\bibinfo{journal}{Nature Photonics}}
  \textbf{\bibinfo{volume}{6}}, \bibinfo{pages}{41--44} (\bibinfo{year}{2011}).

\bibitem{Zhao2009}
\bibinfo{author}{Zhao, B.} \emph{et~al.}
\newblock \bibinfo{title}{{A millisecond quantum memory for scalable quantum
  networks}}.
\newblock \emph{\bibinfo{journal}{Nature Physics}}
  \textbf{\bibinfo{volume}{5}}, \bibinfo{pages}{95--99} (\bibinfo{year}{2009}).

\bibitem{Kaczmarek}
\bibinfo{author}{Kaczmarek, K.~T.} \emph{et~al.}
\newblock \bibinfo{title}{{A room-temperature noise-free quantum memory for
  broadband light}}.
\newblock \emph{\bibinfo{journal}{Preprint at http://arxiv.org/abs/1704.00013}}  (\bibinfo{year}{2017}).

\bibitem{Einstein1935}
\bibinfo{author}{Einstein, A.} \bibinfo{author}{Podolsky, B.} \& \bibinfo{author}{Rosen, N.}
\newblock \bibinfo{title}{{Can Quantum-Mechanical Description of Physical Reality Be Considered Complete?}}
\newblock \emph{\bibinfo{journal}{Physical Review}}
  \textbf{\bibinfo{volume}{47}}, \bibinfo{pages}{777} (\bibinfo{year}{1935}).
  
\bibitem{Hall2011}
\bibinfo{author}{Hall, M.~A.}, \bibinfo{author}{Altepeter, J.~B.} \&
  \bibinfo{author}{Kumar, P.}
\newblock \bibinfo{title}{{Ultrafast Switching of Photonic Entanglement}}.
\newblock \emph{\bibinfo{journal}{Physical Review Letters}}
  \textbf{\bibinfo{volume}{106}}, \bibinfo{pages}{053901}
  (\bibinfo{year}{2011}).

\bibitem{Leszczynski2017}
\bibinfo{author}{Leszczy{\'{n}}ski, A.}, \bibinfo{author}{Parniak, M.} \&
  \bibinfo{author}{Wasilewski, W.}
\newblock \bibinfo{title}{{Phase matching alters spatial multiphoton processes
  in dense atomic ensembles}}.
\newblock \emph{\bibinfo{journal}{Optics Express}}
  \textbf{\bibinfo{volume}{25}}, \bibinfo{pages}{284--295}
  (\bibinfo{year}{2017}).

\bibitem{Hetet2015}
\bibinfo{author}{H{\'{e}}tet, G.} \& \bibinfo{author}{Gu{\'{e}}ry-Odelin, D.}
\newblock \bibinfo{title}{{Spin wave diffraction control and read-out with a
  quantum memory for light}}.
\newblock \emph{\bibinfo{journal}{New Journal of Physics}}
  \textbf{\bibinfo{volume}{17}}, \bibinfo{pages}{073003}
  (\bibinfo{year}{2015}).

\bibitem{Radnaev2010}
\bibinfo{author}{Radnaev, A.~G.} \emph{et~al.}
\newblock \bibinfo{title}{{A quantum memory with telecom-wavelength
  conversion}}.
\newblock \emph{\bibinfo{journal}{Nature Physics}}
  \textbf{\bibinfo{volume}{6}}, \bibinfo{pages}{894--899}
  (\bibinfo{year}{2010}).

\bibitem{Behbood2013}
\bibinfo{author}{Behbood, N.} \emph{et~al.}
\newblock \bibinfo{title}{{Real-time vector field tracking with a cold-atom
  magnetometer}}.
\newblock \emph{\bibinfo{journal}{Applied Physics Letters}}
  \textbf{\bibinfo{volume}{102}}, \bibinfo{pages}{173504}
  (\bibinfo{year}{2013}).

\bibitem{Lipka2016}
\bibinfo{author}{Lipka, M.}, \bibinfo{author}{Parniak, M.} \&
  \bibinfo{author}{Wasilewski, W.}
\newblock \bibinfo{title}{{Optical frequency locked loop for long-term stabilization of broad-line DFB laser frequency difference}}.
\newblock \emph{\bibinfo{journal}{Applied Physics B, 123:238}}  (\bibinfo{year}{2017}).

\bibitem{Chen2006}
\bibinfo{author}{Chen, S.} \emph{et~al.}
\newblock \bibinfo{title}{Deterministic and storable single-photon source based
  on a quantum memory}.
\newblock \emph{\bibinfo{journal}{Physical Review Letters}}
  \textbf{\bibinfo{volume}{97}}, \bibinfo{pages}{173004}
  (\bibinfo{year}{2006}).

\bibitem{Albrecht2015}
\bibinfo{author}{Albrecht, B.}, \bibinfo{author}{Farrera, P.},
  \bibinfo{author}{Heinze, G.}, \bibinfo{author}{Cristiani, M.} \&
  \bibinfo{author}{de~Riedmatten, H.}
\newblock \bibinfo{title}{Controlled rephasing of single collective spin
  excitations in a cold atomic quantum memory}.
\newblock \emph{\bibinfo{journal}{Physical Review Letters}}
  \textbf{\bibinfo{volume}{115}}, \bibinfo{pages}{160501}
  (\bibinfo{year}{2015}).

\end{thebibliography}

\section*{Acknowledgements}
We acknowledge insightful discussions about the magneto-optical trap
with L. Pruvost, G. Campbell, S. Du, G. Roati, M. Zawada and W. Gawlik, assistance of R.
Chrapkiewicz with optical system design, careful proofreading of the
manuscript by K. T. Kaczmarek, M. Jachura, R. Chrapkiewicz, R. \L{}apkiewicz and M. Semczuk and generous
support of K. Banaszek. The project has been funded by National Science
Centre (Poland) grants No. 2015/19/N/ST2/01671, 2016/21/B/ST2/02559
and Polish MNiSW ``Diamentowy Grant'' Projects No. DI2013 011943, DI2016 014846.

\section*{Author Contributions}
M.P., M.D., M.M. and A.L. performed the experiment and analyzed the
data. M.P., M.D. and M.M. wrote the manuscript assisted by other authors.
W.W. managed the project. All authors contributed to building of the
experimental setup.

\section*{Competing interests}
The authors declare no competing financial interests.

\section*{Correspondence}
Correspondence should be addressed to \textsuperscript{\textdagger}M. Parniak at michal.parniak@fuw.edu.pl or to \textsuperscript{*}M. D\k{a}browski at mdabrowski@fuw.edu.pl.

\section*{Data availability}
The data that support the findings of this study are available from M.P. upon reasonable request.

\newpage
\onecolumngrid
\renewcommand\thefigure{\thesection\arabic{figure}}   
\renewcommand{\figurename}{Supplementary Figure}
\setcounter{figure}{0}    
\section*{Supplementary Figures}

\begin{figure}[h]
\includegraphics[scale=0.85]{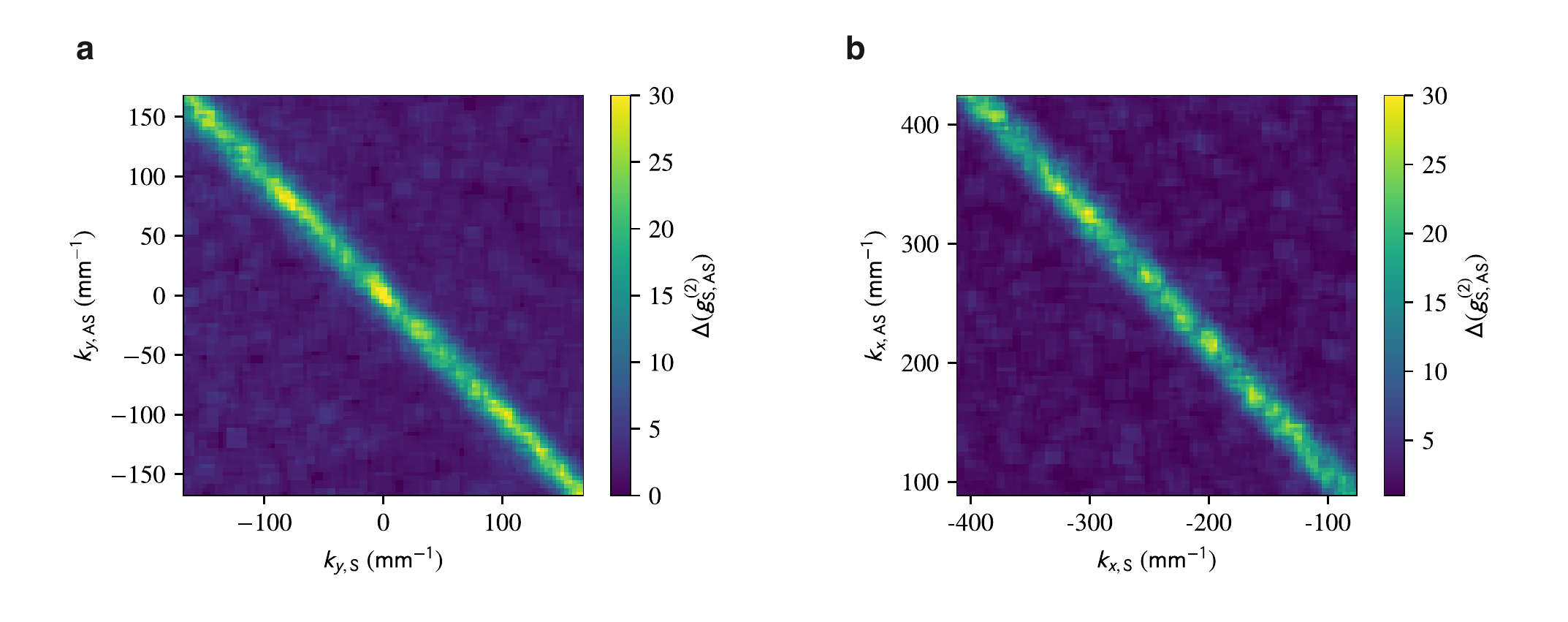}

\caption{\textbf{Standard deviation of results for $g^{(2)}_\mathrm{S,AS}$.} Panels \textbf{(a-b)} correspond to maps from panels a-b in Fig. 4 of the main manuscript. These inferred errors are taken as one standard deviation of a set of results from 25 different sets of regions selected in conjugate columns oriented in $y$-direction ($x$-direction) for a (b).\textbf{\label{fig:mu1}}}
\end{figure}

\begin{figure}[h]
\includegraphics[scale=0.7]{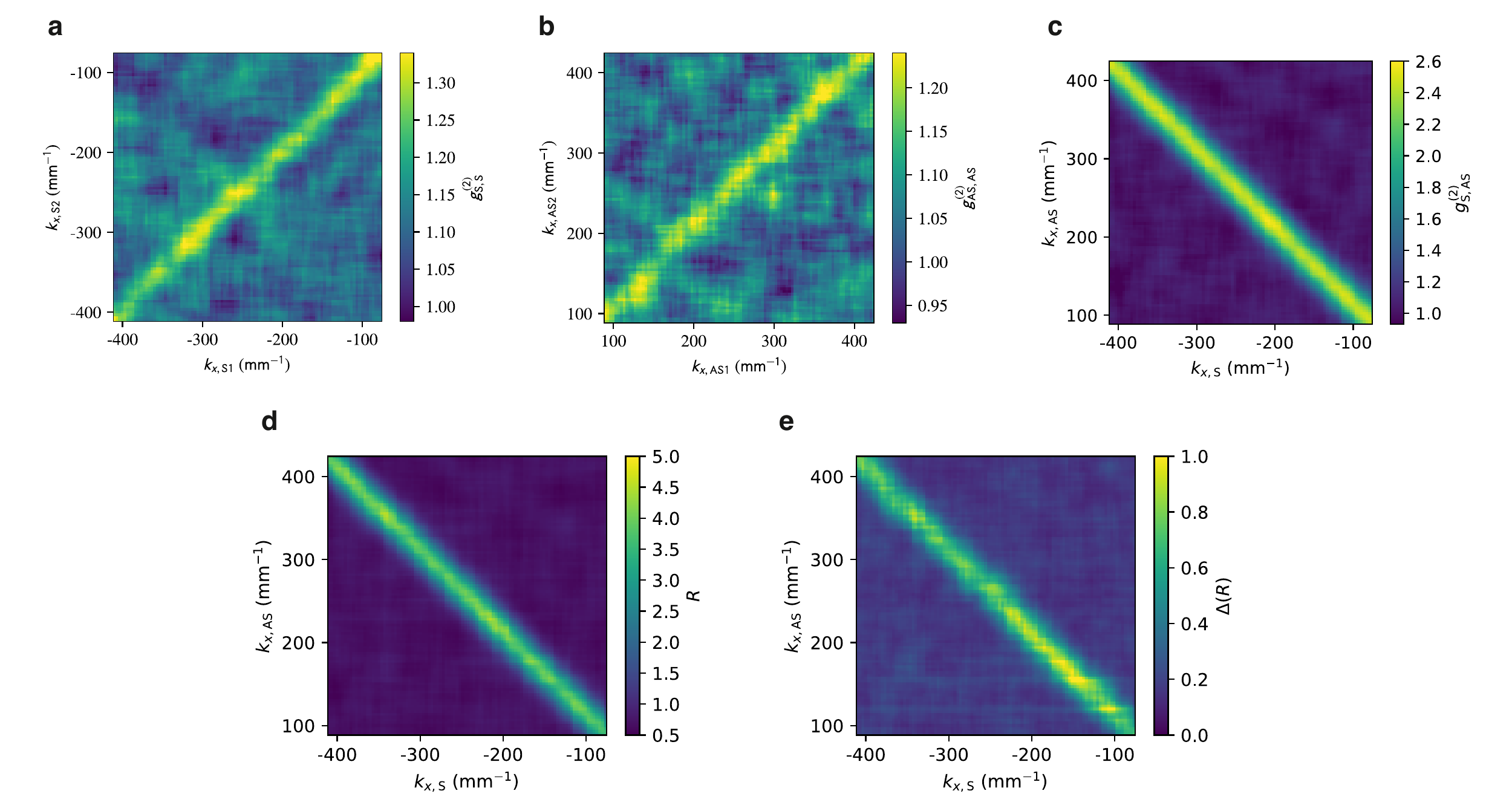}

\caption{\textbf{Results of the autocorrelation measurements.} \textbf{(a-c)} Results of the measurement of $g^{(2)}_\mathrm{S,S}$, $g^{(2)}_\mathrm{AS,AS}$ and $g^{(2)}_\mathrm{S,AS}$, respectively. The measurement was performed using the method described in Methods section of the main manuscript. To obtain better statistics a higher number of generated photons was used than in other measurements, and thus the second-order cross-correlation function is relatively low. Additionally, a larger analysis region with side length $\kappa=25.2\ \mathrm{mm^{-1}}$ was used, resulting in visibly wider diagonal correlation. By averaging diagonal, correlated region we found mean values of  $g^{(2)}_\mathrm{S,S}=1.29\pm0.04$, $g^{(2)}_\mathrm{AS,AS}=1.18\pm0.04$ and $g^{(2)}_\mathrm{S,AS}=2.45\pm0.04$. For a set of uncorrelated regions we found  $g^{(2)}_\mathrm{S,S}=1.07\pm0.04$, $g^{(2)}_\mathrm{AS,AS}=1.04\pm0.03$ and $g^{(2)}_\mathrm{S,AS}=1.02\pm0.05$. These values are consequently higher than the expected value of 1 in the ideal case scenario, which is due to significant classical long-term fluctuations during the measurement.  \textbf{(d)} Inferred value o $R$ demonstrating significant violation of the Cauchy-Schwartz inequality at the diagonal. Respective standard deviations of the values of $R$ are presented in \textbf{(e)}. By averaging the diagonal values we find $R=4.0\pm0.2$. For uncorrelated regions the value is $R=0.68\pm0.06$. Errors correspond to one standard deviation. Analogous results were obtained for the $y$-dimension. \textbf{\label{fig:mu1}}}
\end{figure}

\begin{figure}[h]
\includegraphics[scale=0.85]{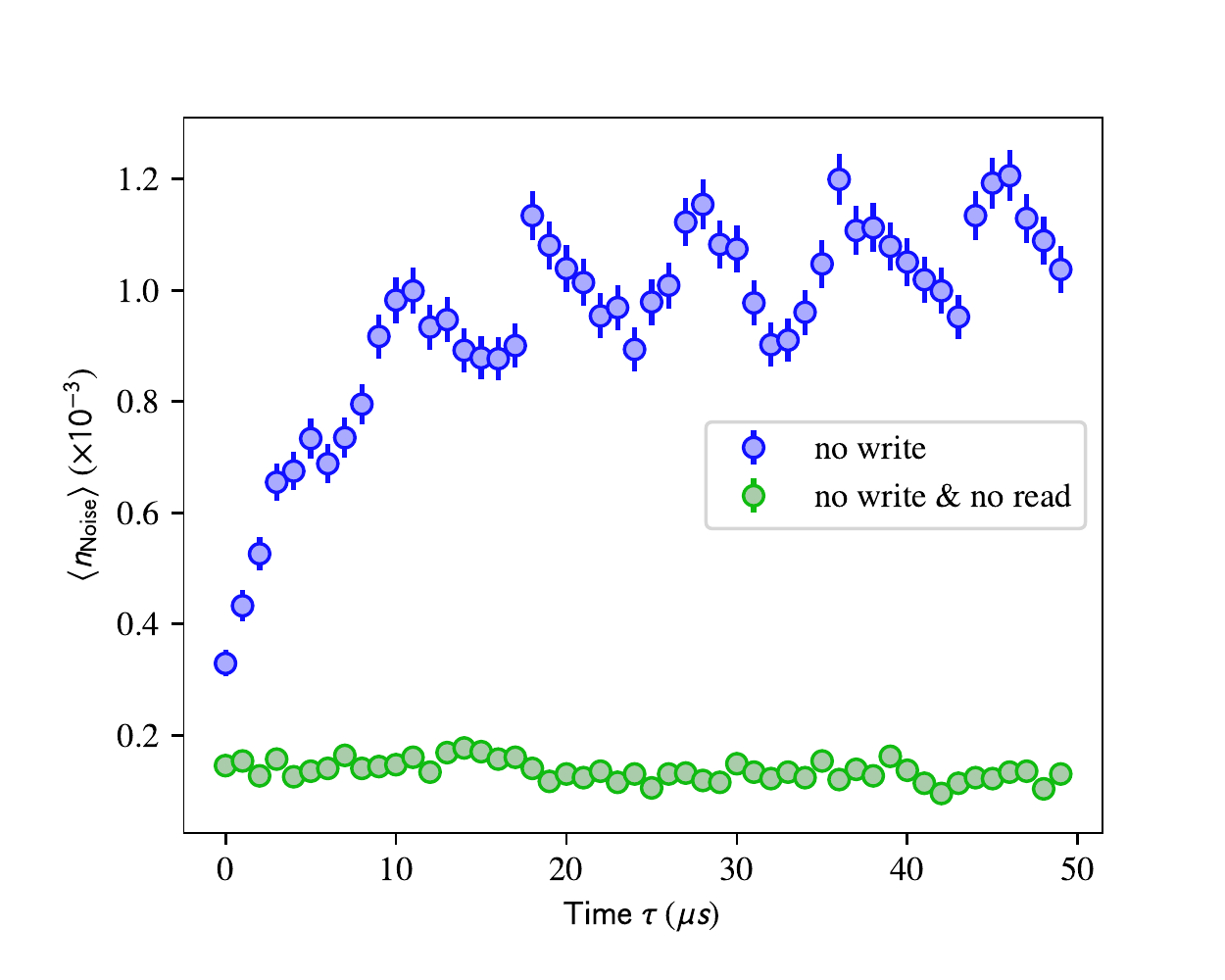}
\caption{\textbf{Noise level as a function of storage time.} The mean number of registered noise photons in the AS arm with (blue dots) and without (green dots) the read laser. The write laser was off in both cases. The noise level without the read laser corresponds to mostly dark counts and residual stray light. With the read laser we observe fluorescence of atoms residing in $F=2$ manifold of the ground state. The initial increase of this fluorescence is due to influx of room-temperature atoms present in the vacuum chamber into the interaction region. The optical pumping ensures that at the zero storage time the fluorescence is minimized. We attribute the oscillation in the noise photons signal to partial spin polarization of room-temperature atoms. Errorbars (one standard deviation) correspond to statistical uncertainty calculated assuming Poissonian distribution of photon counts. Statistical uncertainty for the second case (green circles) is negligible. \textbf{\label{fig:mu1}} }
\end{figure}

\end{document}